\documentclass[aps,prd,twocolumn,nofootinbib,superscriptaddress,longbibliography]{revtex4-1}
\usepackage{graphicx}
\usepackage{tabularx}
\usepackage{mathtools}
\usepackage{amsmath}
\usepackage{amstext}
\usepackage{amssymb}
\usepackage{xfrac}
\usepackage[colorlinks,citecolor=blue]{hyperref}
\usepackage{graphicx}
\usepackage{amsmath}
\usepackage{amstext}
\usepackage{amssymb}
\usepackage{amsfonts}
\usepackage{longtable,booktabs}
\usepackage{hyperref}
\usepackage{url}
\usepackage{subfigure}
\usepackage{dsfont}
\usepackage{booktabs}
\usepackage{amsbsy}
\usepackage{dcolumn}
\usepackage{amsthm}
\usepackage{bm}
\usepackage{esint}
\usepackage{multirow}
\usepackage{hyperref}
\usepackage{cleveref}
\usepackage{mathrsfs}
\usepackage{amsfonts}
\usepackage{amsbsy}
\usepackage{dcolumn}
\usepackage{bm}
\usepackage{multirow}
\usepackage{color}
\usepackage{extarrows}
\usepackage{datetime}
\usepackage{comment}
\usepackage[super]{nth}
\usepackage{tikz-cd}
\usepackage[compat=1.1.0]{tikz-feynman}

\hypersetup{
	colorlinks=magenta,
	linkcolor=blue,
	filecolor=magenta,
	urlcolor=magenta,
}

\newcommand{\bra}{\langle}
\newcommand{\ket}{\rangle}

\begin{document}

	\title{Renormalization group analysis on emergence of higher rank symmetry and  higher moment conservation}
	\author{Hongchao Li}  \altaffiliation{Present address: Department of Physics, University of Tokyo, 7-3-1 Hongo, Bunkyo-ku, Tokyo 113-0033, Japan}
		
	\affiliation{School of Physics and State Key Laboratory of Optoelectronic Materials and
		Technologies, Sun Yat-sen University, Guangzhou, 510275,
		China}
	\affiliation{Department of Modern Physics, University of Science and
		Technology of China, Hefei, 230026, China}
	
	\author{Peng Ye}
	\email{yepeng5@mail.sysu.edu.cn}
	\affiliation{School of Physics and State Key Laboratory of Optoelectronic Materials and
		Technologies, Sun Yat-sen University, Guangzhou, 510275,
		China}
	
	\begin{abstract}
		Higher rank symmetry and higher moment conservation have been drawn considerable attention from, e.g., subdiffusive transport to fracton topological order. In this paper, we perform a one-loop renormalization group (RG) analysis  and show how  these phenomena emerge at low energies.  We consider a $d$-dimensional model of interacting  bosons of $d$ components. At higher-rank-symmetric points with conserved angular moments, the $a$-th bosons have kinetic energy only along the $\hat{x}^a$ direction. Therefore, the symmetric points look highly anisotropic and fine-tuned.  By studying RG in a wide vicinity of the symmetric points, we find that symmetry-disallowed kinetic terms tend to be irrelevant within the perturbative regime, which potentially leads to  emergent higher-rank symmetry and  higher-moment conservation at the deep infrared limit. While   non-perturbative analysis is called for in the future, by regarding higher-rank symmetry as an emergent phenomenon, the RG analysis presented in this paper holds alternative promise for  realizing higher-rank symmetry and higher-moment conservation in experimentally achievable systems.	
	\end{abstract}
	
	\maketitle
	
	\section{Introduction}
	The celebrated Noether's theorem \cite{Noether1918} relates a conservation law to an underlying continuous symmetry.  For example, in a $U(1)$-symmetric Hamiltonian of bosons,  bosonic operator $\hat{\Phi}(\mathbf{x}) $ is changed to $  e^{i\theta}\hat{\Phi}(\mathbf{x})$ under symmetry transformations where  the real parameter $\theta$ doesn't depend on coordinate $\mathbf{x}=(x^1,x^2,\cdots,x^d)$ in $d$-dimensional space.  By means of   Noether's theorem, one can show that the total boson number, i.e., $\int d^d x \rho(\mathbf{x})$, is a conserved quantity, where particle number density  $\rho(\mathbf{x})=\hat{\Phi}^\dagger(\mathbf{x})\hat{\Phi}(\mathbf{x})$.  Apparently, $\rho(\mathbf{x})$ is just the zeroth order  of conventional multipole expansions: 
	\begin{align}  \rho,\,\,\,\, \rho\, \mathbf{x},\,\,\,\,\rho x^a x^b,\,\, \cdots 
	\end{align}
	in  a standard electromagnetism textbook \cite{Griffiths:1492149}. In a particle-number-conserving system, \textit{higher moment conservation}, e.g., conservation of dipoles and quadrupoles, is in principle allowed. Furthermore, if the density is vector-like with multi-components, denoted as $\bm{\rho}=(\rho^1,\rho^2,\cdots,\rho^d)$, then we can define another set of multipole expansions:
	\begin{align}
		\bm{\rho},\,\,\,\, \bm{\rho}\cdot\mathbf{x} ,\,\,\,\,\sum_{e,f=1}^d\epsilon^{ab\cdots ef}\rho^e  x^f,\, \,\cdots  
	\end{align}
	where the third one is \textit{angular moment}. $\epsilon^{ab\cdots}$ is Levi-Civita symbol. If $d=2,3$, it can be rewritten in a compact form: $\bm{\rho}\times \mathbf{x}$. In $d=2$, $\bm{\rho}\times \mathbf{x}=\rho^1 x^2-\rho^2 x^1$.
	
	Indeed,  recently we have been witnesses to  ongoing research progress   on    higher-moment conservation and the associated    higher-rank version of global symmetry \cite{Nandkishore2019,2020Fracton,Pretko2018,Pretko2017a,Pretko2017,PhysRevX.9.031035,2020arXiv200400015S}, especially in the field of  {fracton physics} \cite{Nandkishore2019,2020Fracton,Vijay2015,Vijay2016,Prem2017,Chamon05,Vijay2015,Shirley2019,Ma2017,Haah2011,Bulmash2019,Prem2019,Bulmash2018,Tian2018,You2018,Ma2018,Slagle2017,Halasz2017,Tian2019,Shirley2019b,Shirley2018a,Slagle2019a,Shirley2018,Prem2017,Prem2018,Pai2019,Pai2019a,Sala2019,Kumar2018,Pretko2018,Pretko2017,Ma2018,Pretko2017a,Radzihovsky2019,Dua2019,PhysRevLett.122.076403,haahthesis,PhysRevX.9.031035,2019arXiv190411530Y,2019arXiv190408424S,2019arXiv190404815K, 2019arXiv190913879W, 2019arXiv191213485W,pretko18string,pretko18localization,PhysRevB.100.125150,PhysRevB.99.245135,PhysRevB.97.144106,PhysRevB.99.155118,MaHigherRankDQC, 2019arXiv191101804W,2020arXiv200803852S, 2020arXiv200707894W, 2020arXiv200704904G, 2020arXiv200512317N, 2020arXiv200414393P, 2020arXiv200407251W, 2020arXiv200406115S, 2020arXiv200404181S,2020arXiv200400015S, 2020PhRvR2c3124G, 2020arXiv200212932W, 2020arXiv200212026S, 2020arXiv200205166A, 2020arXiv200202433W, 2020arXiv200105937P,ye19a,Li:2021umc,FS1,FS2}. Some typical examples of research  include subdiffusive transport at late times, non-ergodicity,  Hilbert space fragmentation, and spontaneous symmetry breaking \cite{Sala2019,Pai2019,2019arXiv190404815K,Moudgalya2020,TaylorSR2020,2019arXiv190404815K,moudgalya2019thermalization,PhysRevB.101.125126,PhysRevLett.125.245303,FS1,FS2}.   In a simple scalar theory, the associated higher-rank symmetry transformations are parametrized by $\theta(\mathbf{x})$ that is a polynomial function of  $\mathbf{x}$ \cite{PhysRevX.9.031035}.   
	Inspired by the conventional correspondence between global symmetry and gauge symmetry, upon   ``gauging'' higher-rank symmetry,  higher-rank gauge fields can be obtained \cite{Pretko2018}. Here, the gauge fields are usually higher-rank symmetric tensor fields, which leads to generalized Maxwell equations \cite{Pretko2017a} and exotic theory of  spin systems in Yb-based breathing pyrochlores \cite{Yanprl2020}.

	As a nontrivial consequence of higher moment conservation,  the mobility of particles is inevitably restricted, either partially or completely.  For example, it is quite intuitive that dipole conservation strictly forbids a single particle motion along all spatial directions. Such particles are called ``fractons'' or $0$-dimensional particles \cite{Nandkishore2019,2020Fracton}. Similarly, one can define \textit{lineons} ($1$-dimensional particle) that are movable within a stack of parallel straight lines and \textit{planons} ($2$-dimensional particle) that are movable within a stack of parallel planes. Regarding these strange particles as bosons, we can consider their Bose-Einstein condensation, such that the spontaneous breaking of higher-rank symmetry occurs. As a result, a class of exotic quantum phases of matter dubbed \emph{fractonic  superfluids} \cite{FS1,FS2} is formed. In Ref.~\cite{FS2}, a convenient notation  $d\mathsf{SF}^{i}$ was introduced to denote $d$-dimensional fractonic superfluids with $i$-dimensional particle condensation, e.g., $d\mathsf{SF}^0$ with condensed fractons  and  $d\mathsf{SF}^1$ with condensed lineons. The conventional superfluid phase corresponds to $d\mathsf{SF}^d$ where bosons can freely move.

	Higher-rank symmetric microscopic models often look quite unrealistic, highly anisotropic and fine-tuned \cite{Pretko2018}. For example, Hamiltonian does not has the usual kinetic energy term \cite{FS1}. And interaction is delicately designed \cite{FS2}. However, as we've known, in many condensed matter systems, symmetry  has been found to be {significantly} enhanced  at low energies. For example,  Lorentz symmetry   emerge in graphene which is microscopically built by non-relativistic electrons.  Thus, one may wonder whether it is possible that  long-wavelength low-energy limit  will conserve higher moments and respect higher-rank symmetry as an emergent phenomenon. 
	
	For this purpose, we may apply the traditional theoretical approach: { renormalization group (RG)}. If there  exists a  phase region    such that all models in the region can flow to   symmetric points,  we can regard higher-rank symmetry as  an \textit{emergent  symmetry}.  Theoretically, one advantage of such an emergent higher-rank symmetry is its  robustness against symmetry-breaking perturbation. Practically, we expect that such a scenario holds     promise for \textit{more flexible realization} of exotic higher-rank symmetry and higher-moment conservation in both theoretical and experimental studies.

	In this paper, we identify such a wide phase region that supports emergent higher-rank symmetry and conservation of \textit{angular moments}, i.e., $\int d^2x  \bm{\rho}\times \mathbf{x}= \int d^2x  ({\rho}^1 x^2-{\rho}^2 x^1)$ for a two-component boson fields in two dimensions. We start with a two-dimensional many-boson system in  the normal state (i.e. without lineon condensation) of fractonic superfluids (denoted as $2\mathsf{SF}^1$).  The Hamiltonian is a {symmetric point} in  the parameter space where the {$a$-th ($a=1, 2$)} component bosons only have kinetic terms along $a$-th axis (dubbed ``\textit{diagonal}'' kinetic terms).  There also exists a weak inter-component scattering term allowed by higher-rank symmetry.   We shall perform a  RG analysis in the vicinity of the symmetric point by adding symmetry-disallowed kinetic terms (dubbed ``\textit{off-diagonal}'' kinetic terms) as a perturbation. The one-loop calculation of $\beta$-function shows that there exists a finite phase region (Fig.~\ref{figure_RGflow}) where off-diagonal kinetic terms tend to be irrelevant under RG iteration. {In other words, the high-energy model, which is not symmetric but more realistic and less fine-tuned,  has a tendency to flow to the  symmetric point. As a result, higher-rank symmetry as well as conservation of angular moments emerges. 
		
		The remainder of this paper is organized as follows. In Sec.~\ref{section_model}, we introduce the $d$-component bosonic systems and its higher-rank symmetry. In Sec.~\ref{renormalization}, we discuss the scaling and Feynamn rules of the $d$-component bosonic systems. Further, we figure out the $\beta$-functions of parameters in the systems with renormalization group (RG) analysis and depict the global phase diagram. In Sec.~\ref{summary}, we summarize and provide our prediction on conditions of possible realization of the systems with higher-rank symmetry.

		\section{Model and symmetry}\label{section_model}
		The symmetric point  Hamiltonian for the $d$-component bosonic systems in   real space \cite{FS2} is given by $\mathcal{H}=\mathcal{H}_0+\mathcal{H}_1$ with
		\begin{align}
			\!	\!\!\!	\!\mathcal{H}_0\! & =\!\sum_{a = 1}^d \left[\frac{t_a}{2} (\partial_a\hat{\Phi}_a^{\dagger}) (\partial_a \hat{\Phi}_a) - \mu \hat{\Phi}_a^{\dagger}\hat{\Phi}_a\right]\,,\\
			\!	\!\!\!		\mathcal{H}_1 \!& =\!\frac{1}{2} \!\sum_{a \neq b}\! K_{a b} (\hat{\Phi}_a^{\dagger}\partial_a \hat{\Phi}_b^{\dagger}\!+\!\hat{\Phi}_b^{\dagger}  \partial_b \hat{\Phi}_a^{\dagger} )(\hat{\Phi}_a  \partial_a \hat{\Phi}_b \! +\!	\hat{\Phi}_b  \partial_b \hat{\Phi}_a )\,,\!\!\label{eqn_h1}
		\end{align}
		{where $\mathcal{H}_0$ is free Hamiltonian and $\mathcal{H}_1$ is interaction Hamiltonian.} Here $t_a=m^{-1}_a$ stands for the inverse of mass along the $a$-th direction {and $\mu$ stands for the chemical potential}.    $\hat{\Phi}^{\dagger}_a(\mathbf{x})$ and $\hat{\Phi}_a(\mathbf{x})$   are respectively creation and annihilation operators of $d$-component bosons, and satisfy the bosonic commutation relations. The interaction strength $K_{ab}$ is a symmetric matrix with vanishing diagonal elements, i.e., $K_{aa}=0, K_{ab}=K_{ba}$.  Each term in $\mathcal{H}$ is invariant under both the conventional global symmetry transformations ($\hat{\Phi}_a \rightarrow e^{i\theta_a}\hat{\Phi}_a$, $\theta_a\in\mathbb{R}$) and higher-rank symmetry transformations:  
		\begin{align}
			(\hat{\Phi}_a,\hat{\Phi}_b)\longrightarrow(\hat{\Phi}_ae^{i\lambda_{ab}x^b},\hat{\Phi}_be^{i\lambda_{ba}x^a})\label{eq_symmetry}
		\end{align} 
		for each pair $(\phi_a,\phi_b)$ with $\lambda_{ab}=-\lambda_{ba}\in\mathbb{R}$. { According to the Noether's theorem, the conventional global $U(1)$ symmetry and the higher-rank symmetry correspond to conserved total charge (particle number) $\hat{\mathcal{Q}}^a$ and conserved total angular moments $\hat{\mathcal{Q}}^{ab}$ ($\hat{\mathcal{Q}}^{ab}=-\hat{\mathcal{Q}}^{ba}$) \cite{Noether1918,Pretko2018,FS2}:
			\begin{align}
				\hat{\mathcal{Q}}^a=&\int d^dx\hat{\rho}^a\,,\,	\,	\,	\hat{\mathcal{Q}}^{ab}=\int d^dx(\hat{\rho}^a x^b-\hat{\rho}^b x^a)\,.\label{conserved_angular_total}
			\end{align}
			Here  $\hat{\rho}^a=\hat{\Phi}^{\dagger}_a\hat{\Phi}_a$. Intuitively, the conserved quantities $\mathcal{Q}^{ab}$ enforce that a single $a$-th component boson  can only move along the $a$-th direction.  More explanation on the classical mechanical consequence of the conservation is available  in  Appendix~\ref{appendix_a} and Ref.~\cite{FS2}}.

		In the coherent-state path-integral formulation with imaginary time, the Lagrangian density $\mathcal{L}$ can be written as $\mathcal{L}=\phi_a^{\ast} \partial_{\tau} \phi_a+\mathcal{H}$ with action $S = \int d \tau d^d x \mathcal{L}$. Here the bosonic fields $\phi_a=\phi_a(\mathbf{x},\tau)\in\mathbb{C}$ are the eigenvales of coherent-state operators $\hat{\Phi}_a(\mathbf{x},\tau)$.  
		With the Fourier transformation of the coherent state: $\phi_a(\tau,\mathbf{r})=\frac{1}{\sqrt{\beta}}\sum_{n}\int\frac{d^dk}{(2\pi)^d}e^{i\mathbf{k}\cdot\mathbf{r}-i\omega_n\tau}\phi_a(i\omega_n,\mathbf{k})$ and its complex conjugate: $\phi_a^*(\tau,\mathbf{r})=\frac{1}{\sqrt{\beta}}\sum_{n}\int\frac{d^dk}{(2\pi)^d}e^{-i\mathbf{k}\cdot\mathbf{r}+i\omega_n\tau}\phi_a^*(i\omega_n,\mathbf{k})$ for $a=1,2,\cdots,d$ { with $\beta=\frac{1}{k_BT}$ where $T$ represents the temperature of bosons}.   In the frequency-momentum space, 
		    \begin{align}	
				S  = & \sum_{a;k} (-i\omega_n+\xi_a) \phi_a^{\ast}  \phi_a   
				+ \sum_{\substack{1,2,3,4\\{a\neq b}}} \frac{K_{a b}}{2\beta}  (k_2^a + k_1^b) ( {k}_{3}^{a} + k_{4}^{b})
				\nonumber\\
				&\cdot \phi_{1 a}^{\ast} \phi_{2 b}^{\ast} \phi_{3 b} \phi_{4 a} \delta (1 + 2 - 3 - 4).\label{equation_action_Main}
			\end{align}  
			In the first term on the r.h.s.,  the simplified notation $\phi_a$ stands for $\phi_a (i \omega_{n}, \mathbf{k})$ which is the frequency-momentum image of the field $\phi_a(\tau, \mathbf{r})$.  $\omega_{n}$ is a bosonic Matsubara frequency and $\mathbf{k}$ is a momentum vector: $\mathbf{k}=({k}^1,{k}^2,\cdots,{k}^d)$. We use $k^a$ to denote the $a$-th spatial component of momentum vector $\mathbf{k}$.   $\sum_{k}$ stands for $\sum_{\omega_{n}}\int\frac{d^d  {k}}{(2\pi)^d}$. In the second term, since there are four pairs of frequency and momentum,   we introduce a new notation $\phi_{ia}$  to compactly represent  $\phi_{a}( i\omega_{n_i},\mathbf{k}_{i})$ where the label $i=1,2,\cdots,4$.  $k_i^a$ stands for the $a$-th spatial component of momentum vector $\mathbf{k}_i$.     The sum $\sum_{1,2,3,4}$  denotes  $\sum_{i=1}^4\sum_{\omega_{n_i}}\int\frac{d^d  {k_i}}{(2\pi)^d}$. Other notations like $\sum_{1,2},\sum_{2},\dots$ in the forthcoming texts are defined in the similar way. 
			{Besides, the kinetic energy with momentum $\mathbf{k}$ is defined as: $\xi_{a}=\frac{1}{2}{t_a(k^{a})^2}-\mu$(In the following text, we began to introduce anisotropic kinetic energy)}.   For momentum $\mathbf{k}_i$, the associated kinetic energy is $\xi_{ia}=\frac{1}{2}\sum_{b=1}^d t_{ab}(k_{i}^b)^2 -\mu$.      Last, we use $\delta(1+2-3-4)$ to represent $\delta_{(n_1+n_2), (n_3+n_4)}(2\pi)^d\delta(\mathbf{k_1}+\mathbf{k_2}-\mathbf{k_3}-\mathbf{k_4})$. $\delta_{(n_1+n_2), (n_3+n_4)}$ is a dimensionless Kronecker symbol.}
		
		Before moving forward, we perturb the Lagrangian density by adding small ``off-diagonal'' kinetic terms that break higher-rank symmetry. {They are the terms deviating the model from the symmetric point}. As such, kinetic terms of both directions are present, which can be written as  
		$\sum_{a,b = 1}^d \left[ \frac{t_{a b}}{2} (\partial_b \phi_a^{\ast})
		(\partial_b \phi_a) \right]=\sum_{a=1}^d \left[ \frac{t_{a a}}{2} (\partial_a \phi_a^{\ast})
		(\partial_a \phi_a) \right]+\sum_{a \neq  b}^d \left[ \frac{t_{a b}}{2} (\partial_b \phi_a^{\ast})
		(\partial_b \phi_a) \right]$. 
		The kinetic parameter $t_a$ is rewritten as $t_{a a}$ for the notational convenience. Those off-diagonal kinetic terms with nonzero $t_{ab}$ ($a\neq b$) manifestly break higher-rank symmetry. Similarly, we can also understand these parameters as inverse of mass of field configuration $\phi_a$ along other directions other than the $a$-th one: $t_{ab}=1/{m_{ab}}$.{In this way, the kinetic energy with momentum $\mathbf{k}$ can be redefined as: $\xi_{a}=\frac{1}{2}{\sum_{a'=1}^2 t_{aa'}(k^{a'})^2}-\mu$.}
		
		\begin{figure}[t]
			\centering
			\includegraphics[width=0.4\columnwidth]{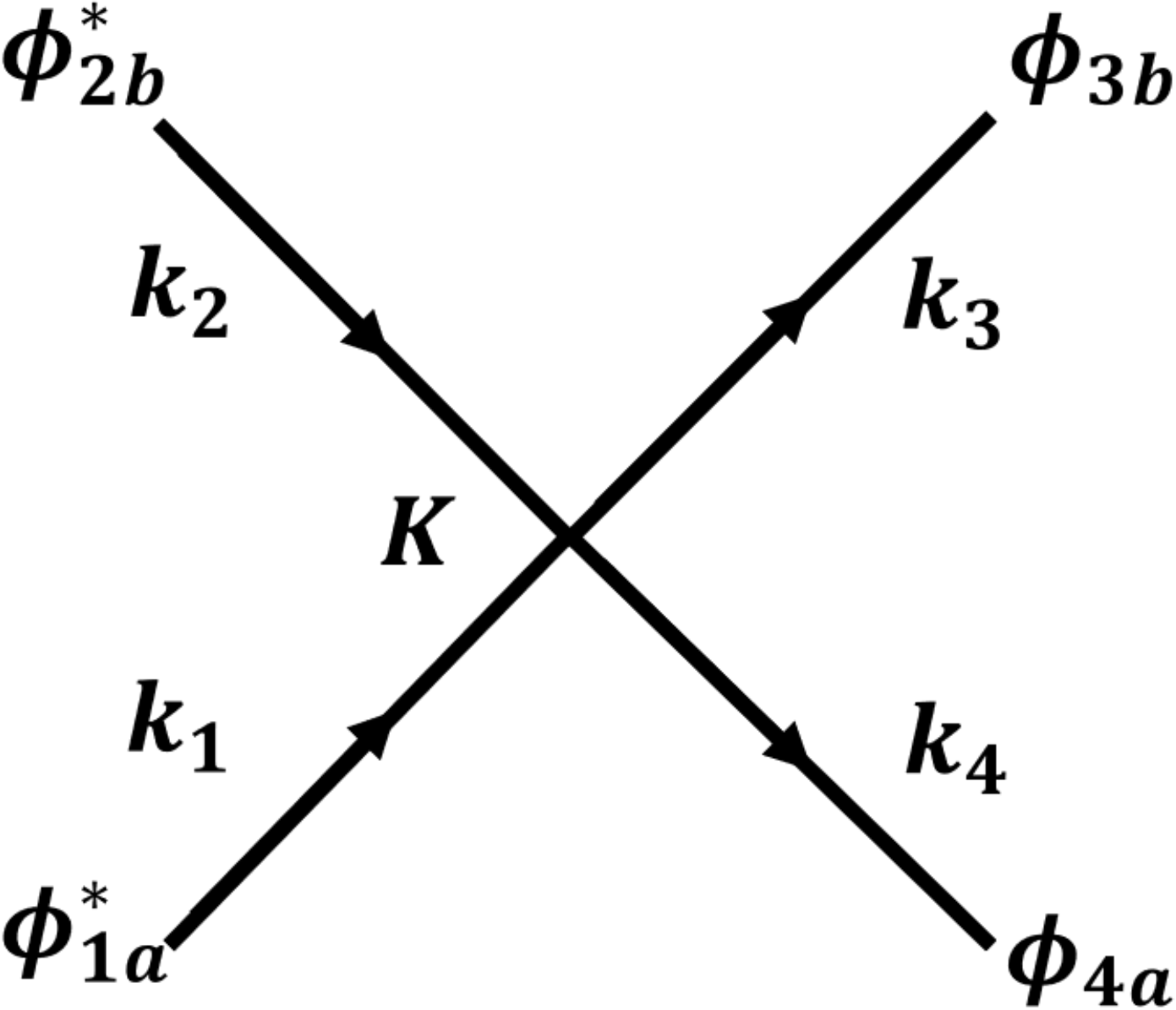}\phantom{AAA}
			\includegraphics[width=0.45\columnwidth]{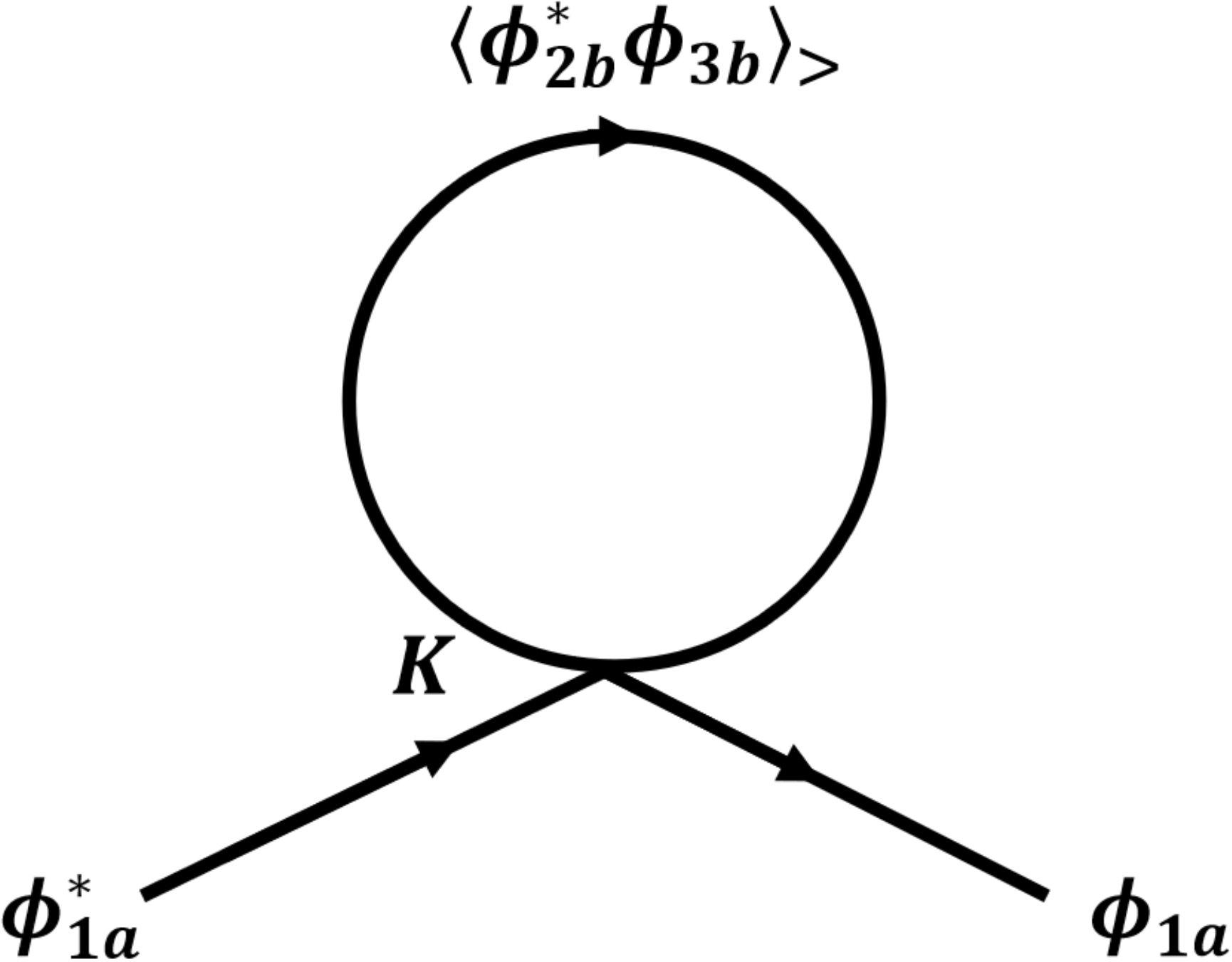}\\
			\caption{The left one shows the bare interaction vertex of $d$-component bosons. Here the solid lines are Feynman propagators and the vertex is the coefficient: $\frac{1}{2\beta}K_{a b} (k_{2}^{a} + k_{1}^{b}) (k_{3}^{a} + k_{4}^{b})\delta (1 + 2 - 3 - 4)$. The right one represents the Feynman loop diagram of the correction for the parameter $\mathcal{T}_1$.}
			\label{fig2}
		\end{figure}
		\section{Renormalization group analysis}\label{renormalization}
		\subsection{Scaling and Feynman rules}
		
		We consider $d=2$. We   set the restriction of field configuration $\phi_a$ as: $\sum_{b=1}^2\frac{t_{ab}}{t_a}(k^b)^2\leq \Lambda^2$ since our total kinetic energy is given by $\sum_b \frac{t_{ab}}{2}(k^b)^2$ with surfaces of equal energy: $\sum_b \frac{t_{ab}}{t_{aa}}(k^b)^2=\kappa^2$.  Here $\kappa$ is an arbitrary constant with the momentum dimension {and $\Lambda$ is the cut-off of momentum here}.  The high-energy part corresponds to: $\sqrt{\sum_{b=1}^2\frac{t_{ab}}{t_a}(k^b)^2}\in[\Lambda/ s, \Lambda]$, where the scaling parameter $s>1$ and sends  $\mathbf{k}$ to $\mathbf{k}/ s $.   We also define $s=e^l$ with $l>0$.  We consider the free part of the action  $ \sum_n\int\frac{d^2 k}{(2 \pi)^2} \phi_a^{*} \!\left( -i \omega_n + \sum_{b=1}^2 \frac{t_{ab}(k^b)^2}{2} \right)\! \phi_a
		$. {In the following text, we also call it fast modes and denote it with $>$.} It is noticed that we do not take the part related to chemical potential into account since it must be relevant if we choose the kinetic part to be marginal. Suppose the scaling dimension of $\phi$ is $\Delta_{\phi}$, and we can assume
		the change of temperature and frequencies are described as:
		$	T \rightarrow s^{- z} T
		$, and $\omega_n\rightarrow s^{-z} \omega_n, n \in
		\mathbb{Z}
		$ respectively. So the free part is scaled to $  \sum_n\!\!\int\!\!\frac{d^2 k}{(2 \pi)^2} \phi_a^{*}\! \left( - i \omega_n
		s^{- z} \!+\! \sum_{b=1}^2 \frac{t_{ab}(k^b)^2}{2} s^{- 2} \right)\! \!\phi_a s^{2 \Delta_{\phi} - 2}.
		$ To fix the momentum dependence, we need
		$	\Delta_{\phi} = \frac{d}{2} + 1=2
		$ and $z = 2$.  
		In $d=2$ case considered here,   the interaction matrix is simply determined by a single parameter, i.e., $K_{12}=K_{21}:= \mathcal{K}$. For further simplifying calculation for $d=2$, we assume $t_1:=t_{11}=t_2:=t_{22}:=\mathcal{T}_0, t_{12}=t_{21}:=\mathcal{T}_1$. { There are two small parameters compared to $\mathcal{T}_0$. The first one is symmetry-disallowed off-diagonal kinetic parameter $\mathcal{T}_1$ which is marginal. It can also been seen as a perturbation relative to the interaction parameter $\mathcal{K}\Lambda^2$. The second one is the irrelevant interaction parameter $\mathcal{K}\Lambda^2$ considered as an infinitesimal quantity compared with diagonal kinetic parameter $\mathcal{T}_0$. Hence, the following calculation will be proceeded in the perturbative regime: $\mathcal{T}_0 \gg \mathcal{K}\Lambda^{2}\gg \mathcal{T}_1$, where $\Lambda$ is the momentum cut-off.}
		
		Next, we write Feynman rules. The bare Feynman propagators are given by:  $\langle \phi^{\ast}_{i a} \phi_{j b} \rangle = (i \omega_{n_i} -
		\xi_{i a})^{-1} \delta_{{ab}} \delta_{{ij}}$ {with path-integral quantization}.   Here $\xi_{ia}=\frac{\sum^d_{a'=1} t_{aa'}(k_{i}^{a'})^2}{2}-\mu$ stands for the kinetic energy of $\phi_{ia}$ [see definition below Eq.~(\ref{equation_action_Main})]. In addition, the interaction vertex is drawn in Fig.~\ref{fig2}.  
		This diagram represents the interaction term, i.e. the second term in Eq.~(\ref{equation_action_Main}), where solid lines represent the bosonic fields and vertex stands for the interaction coefficient.
		In the RG procedure to be performed, we apply  the standard cumulant expansion and relate the mean of the exponential to the exponential of the means \cite{Shankar_1994,Komargodski_2011}: $
		\bra e^{-S_{int}}\ket_>=e^{-\bra S_{int}\ket_>+\frac{1}{2}(\bra S_{int}^2\ket_>-\bra S_{int}\ket_>^2)+...} $. Here $\phi_>$ and $\phi_<$ respectively correspond to the fast modes and slow modes of bosonic fields {and $S_{int}$ stands for interaction part of action}. In this case, the notation $\bra\phantom{A}\ket_>$ is to take the average over the fast modes. Therefore, after calculating the average on fast modes, we have the form of the effective action.
		$ 	S_{eff}[\phi_<] =  S_0[\phi_<]+S_{int}'[\phi_<] 
		=   S_0[\phi_<]+\bra S_{int}\ket_>[\phi_<]-\frac{1}{2}(\bra S_{int}^2\ket_>-\bra S_{int}\ket_>^2)[\phi_<] $.  The form of $\bra S_{int}\ket_>$ contributes to the $\beta$-functions of kinetic parameters, also named first-order correction. Further, $\frac{1}{2}(\bra S_{int}^2\ket_>-\bra S_{int}\ket_>^2)[\phi_<]$ leads to the $\beta$-functions of the elements in the $K$ matrix.

		\subsection{One-loop correction to $\mathcal{T}_1$}	
		The  interaction part of action reads:
		\begin{align}
			S_{int} =& \frac{1}{2\beta} \sum_{1,2,3,4}\sum_{a \neq b} K_{ab} (k_2^a + k_1^b) (k_3^a + k_4^b) \phi_{1a}^{\ast} \phi_{2 b}^{\ast} \phi_{3 b} \phi_{4 a} \nonumber\\
			&\delta (1 + 2 - 3 - 4)\,.
			\label{int} 
		\end{align}
		Below we determine the scaling of the $K$ matrix.  To be specific, we define the new momentum and frequency as: $\mathbf{k}'=s\mathbf{k}$, $\omega_n'=s^z\omega_n$ or $T'=s^zT$ with $z=2$. We define the rescaled field: $\phi_a'(i\omega_n',\mathbf{k}')=s^{-\Delta_{\phi}}\phi_a(i\omega_n's^{-z},\frac{\mathbf{k}'}{s})$ with $\Delta_\phi=\frac{d}{2}+1$.  Then the scaling of parameters $K_{ab}$ can be determined. With the form of interacting action (\ref{int}), we obtain:
		\begin{widetext}
			\begin{align}
				S'_{int}  =&\frac{1}{2\beta'}s^{-z}\sum_{i=1}^4\sum_{\omega_{n_i}'}\int\frac{d^d  {k_i'}}{(2\pi)^d}s^{-4d}K_{ab}((k_2')^a + (k_1')^b) ((k_3')^a + (k_4')^b)s^{-2}\nonumber\\
				&\cdot\phi_{a}'^{\ast}(i\omega_{n_1}',\mathbf{k}_1') \phi_{b}'^{\ast}(i\omega_{n_2}',\mathbf{k}_2') \phi_{b}'(i\omega_{n_3}',\mathbf{k}_3') \phi_{a}'(i\omega_{n_4}',\mathbf{k}_4')s^{4\Delta_{\phi}} \delta_{(n_1'+n_2'),(n_3'+n_4')}(2\pi)^d\delta(\mathbf{k_1'}+\mathbf{k_2'}-\mathbf{k_3'}-\mathbf{k_4'})s^d\nonumber\\
				=&  \frac{1}{2\beta'}s^{2-z-d} \sum_{i=1}^4\sum_{\omega_{n_i}'}\int\frac{d^d  {k_i'}}{(2\pi)^d}\sum_{a \neq b} K_{ab} ((k_2')^a + (k_1')^b) ((k_3')^a + (k_4')^b) \nonumber\\
				&\cdot\phi_{a}'^{\ast}(i\omega_{n_1}',\mathbf{k}_1') \phi_{b}'^{\ast}(i\omega_{n_2}',\mathbf{k}_2') \phi_{b}'(i\omega_{n_3}',\mathbf{k}_3') \phi_{a}'(i\omega_{n_4}',\mathbf{k}_4')\delta_{(n_1'+n_2'),(n_3'+n_4')}(2\pi)^d\delta(\mathbf{k_1'}+\mathbf{k_2'}-\mathbf{k_3'}-\mathbf{k_4'})\,.
				\label{K}
			\end{align}
		\end{widetext}
		Here the dimensionful $\delta$-function of momenta is also scaled [$\delta (\mathbf{k}')=s^{-d}\delta(\mathbf{k})$] but the Kronecker symbol of Matsubara frequencies is dimensionless and invariant upon scaling. 
		Therefore, the scaling of $K_{ab}$ is given by $K'_{ab}=K_{ab}s^{-d}$.   It illustrates that the $K$ matrix at tree level is irrelevant in perturbative RG.

		Then, let us consider the first-order correction to $\mathcal{T}_1$: $\bra S_{int}\ket_{>}$ contributing to the correction of kinetic parameters.
		\begin{align}
			\bra S_{int}\ket_{>} = &\bigg\bra \frac{1}{2\beta} \sum_{1,2,3,4}\sum_{a \neq b} K_{ab} (k_2^a + k_1^b) (k_3^a + k_4^b) \nonumber\\
			&\cdot\phi_{1a}^{\ast} \phi_{2 b}^{\ast} \phi_{3 b} \phi_{4 a} \delta (1 + 2 - 3 - 4)\bigg\ket_{>}\,.
		\end{align}
		To proceed further, we should split each momentum integration into slow part ($<$) and fast part ($>$). 
		In this way, the integration of four momenta $(\mathbf{k}_1,\mathbf{k}_2,\mathbf{k}_3,\mathbf{k}_4)$ is split to $2^4=16$ combinations. 
		According to Wick's theorem, for formulas of bare propagators, we conclude that either $\phi^*_{1a}$ and $\phi_{4a}$ must be paired or $\phi^*_{2b}$ and $\phi_{3b}$ must be paired. Other contractions vanish due to $K_{ab}=0$ when $a=b$. 
		Therefore, in $16$ combinations, only the following two are  non-vanishing:
		\begin{itemize}
			\item  \textbf{Case-I:} The momenta carried by $\phi_{1a}^*$ and $\phi_{4a}$, i.e., $\mathbf{k}_1$ and $\mathbf{k}_4$, are fast momenta\footnote{By fast (slow) momentum, we mean that the kinetic energy corresponding to the momentum is large (small).}. $\mathbf{k}_1=\mathbf{k}_4$ and $\omega_{n_1}=\omega_{n_4}$  required by the formula of bare propagator;
			\item  \textbf{Case-II:} The momenta carried by $\phi_{2b}^*$ and $\phi_{3b}$, i.e., $\mathbf{k}_2$ and $\mathbf{k}_3$, are fast momenta. $\mathbf{k}_2=\mathbf{k}_3$ and $\omega_{n_2}=\omega_{n_3}$ required by the formula of bare propagator.
		\end{itemize}
		By further considering $\delta(1+2-3-4)$, we have:
		\begin{itemize}
			\item   \textbf{Case-I:} The momenta carried by $\phi^*_{1a}$ and $\phi_{4a}$, i.e., $\mathbf{k}_1$ and $\mathbf{k}_4$, are fast momenta. $\mathbf{k}_1=\mathbf{k}_4$, $\omega_{n_1}=\omega_{n_4}$, $\mathbf{k}_2=\mathbf{k}_3$, and $\omega_{n_2}=\omega_{n_3}$;
			\item  \textbf{Case-II:} The momenta carried by $\phi^*_{2b}$ and $\phi_{3b}$, i.e., $\mathbf{k}_2$ and $\mathbf{k}_3$, are fast momenta. $\mathbf{k}_2=\mathbf{k}_3$ and $\omega_{n_2}=\omega_{n_3}$, $\mathbf{k}_1=\mathbf{k}_4$, and $\omega_{n_1}=\omega_{n_4}$.
		\end{itemize}	
		These two contractions correspond to the Feynman diagram in Fig.~\ref{fig2}. We firstly focus on case-I. When 1,4 are fast modes, the momentum conservation and Feynman rules will tell us: $\mathbf{k}_2 =\mathbf{k}_3, \mathbf{k}_1=\mathbf{k}_4 ; \omega_{n_2} = \omega_{n_3}, \omega_{n_1} = \omega_{n_4}$. Since we focus on  the two-dimensional case $(d=2)$, we set all off-diagonal $K$-matrix elements as: $K_{12}=K_{21}:= \mathcal{K}$. For further simplifying calculation, we assume $t_1:=t_{11}=t_2:=t_{22}:=\mathcal{T}_0$, $t_{12}=t_{21}:=\mathcal{T}_1$.  
		In this way, $\langle S_{{int}} \rangle_{>}^{I}$ is given by:
		\begin{align}
			\langle S_{int} \rangle_{>}^{I} = & \frac{1}{2\beta} \sum_{a \neq b}\sum_{1,2} \mathcal{K} (k_2^a + k_1^b)^2 \langle \phi^{\ast}_{1 a} \phi_{1 a}\rangle_{>} \phi_{2 b}^{\ast} \phi_{2 b} \nonumber\\
			=&  \frac{1}{2\beta} \sum_{a \neq b} \sum_{1,2} \mathcal{K} (k_2^a + k_1^b)^2\frac{1}{i \omega_{n_1} - \xi_{1 a}} \phi_{2 b}^{\ast} \phi_{2 b}\nonumber\\
			= & -\frac{1}{2} \sum_{a \neq b} \sum_2  \int_{>} \frac{d^2 k_1}{(2\pi)^2} \mathcal{K} (k_2^a + k_1^b)^2 f_B (\xi_{1 a}) \phi_{2 b}^{\ast}\phi_{2 b} \nonumber\\
			= & -\frac{1}{2} \sum_{a \neq b} \sum_2 \int_{>} \frac{d^2 k_1}{(2\pi)^2} \mathcal{K} \left[(k_2^a)^2 + 2 k_2^a k_1^b + (k_1^b)^2\right] \nonumber\\
			&\cdot f_B (\xi_{1 a})\phi_{2 b}^{\ast} \phi_{2 b} \,,
		\end{align} 
		where $\mathbf{k}_1$ and $\mathbf{k}_2$ correspond to fast and slow momenta respectively.  The bosonic Matsubara summation is applied:
		$		\frac{1}{\beta}\sum_{\omega_n}\frac{1}{i\omega_{n}-\xi_{1a}}=-f_{B}(\xi_{1a})=-\frac{1}{e^{\beta\xi_{1a}}-1}\,.
		$ And the propagator $\langle \phi^{\ast}_{1 a} \phi_{1 a}\rangle_{>} $ is given by:
		$
		\langle \phi^{\ast}_{1 a} \phi_{1 a}\rangle_{>} =\frac{1}{i \omega_{n_1} - \xi_{1 a}} 
		$ with $\xi_{1a}$   ($a=1, b=2$ or $a=2,b=1$):
		\begin{align}
			\xi_{1a} :=
			\frac{\mathcal{T}_0 (k_1^a)^2+\mathcal{T}_1 (k_1^b)^2}{2}-\mu\,. \label{eq_dispersion}
		\end{align}

		In the same way, we can give the form of  case-II:
		\begin{align}
			&	\langle S_{int} \rangle_{>}^{II} =  \frac{1}{2\beta} \sum_{a \neq b}\sum_{1,2} \mathcal{K} (k_2^a + k_1^b)^2 \langle \phi^{\ast}_{2b} \phi_{2b}\rangle_{>} \phi_{1a}^{\ast} \phi_{1a} \nonumber\\
			= & \frac{1}{2\beta} \sum_{a \neq b} \sum_{1,2} \mathcal{K} (k_2^a + k_1^b)^2\frac{1}{i \omega_{n_2} - \xi_{2b}} \phi_{1a}^{\ast} \phi_{1a}\nonumber\\
			= & -\frac{1}{2} \sum_{a \neq b} \sum_1 \int_> \frac{d^2 k_2}{(2\pi)^2} \mathcal{K} (k_2^a + k_1^b)^2 f_B (\xi_{2b}) \phi_{1a}^{\ast}\phi_{1a} \nonumber\\
			= & -\frac{1}{2} \sum_{a \neq b} \sum_1 \int_> \frac{d^2 k_2}{(2\pi)^2} \mathcal{K} \left[(k_2^a)^2 + 2 k_2^a k_1^b + (k_1^b)^2) \right]\nonumber\\
			&\cdot f_B (\xi_{2b})\phi_{1a}^{\ast} \phi_{1a} \,.\nonumber
		\end{align}
		By observing these two actions, we find they are completely equivalent to  each other by exchanging the indices ($a\leftrightarrow b\,, 1\leftrightarrow 2$). Then we focus on the case-I and multiply it by $2$, namely:
		\begin{align}
			&\langle S_{int} \rangle_{>}  = \langle S_{int} \rangle_{>}^{I}+\langle S_{int} \rangle_{>}^{II}=2\langle S_{int} \rangle_{>}^{I}\,,\nonumber\\
			& = -\sum_{a \neq b} \sum_2 \int_{>} \frac{d^2 k_1}{(2\pi)^2} \mathcal{K}  \left[(k_2^a)^2 + 2 k_2^a k_1^b + (k_1^b)^2) \right]\nonumber\\
			&\cdot f_B (\xi_{1 a})\phi_{2 b}^{\ast} \phi_{2 b} = \langle S_{int} \rangle_{>}^{(1)}+\langle S_{int} \rangle_{>}^{(2)}+\langle S_{int} \rangle_{>}^{(3)}\,,\nonumber
		\end{align}
		where we arrange all terms into three parts:
		\begin{align}
			\langle S_{{int}} \rangle_{>}^{(1)}:=&-\sum_{a \neq b} \sum_2 \mathcal{K} (k_2^a)^2 \int_> \frac{d^2 k_1}{(2 \pi)^2} f_B (\xi_{1a}) \phi_{2 b}^{\ast} \phi_{2 b}\,,\label{part_1}\\
			\bra S_{int}\ket_{>}^{(2)}:=&-2\sum_{a \neq b} \sum_2 \mathcal{K} k_{2}^a\int_> \frac{d^2 k_1}{(2\pi)^2}  k_{1}^b f_B (\xi_{1a})\phi_{2b}^{\ast} \phi_{2b}\,,\label{part_2}\\
			\langle S_{{int}} \rangle_{>}^{(3)}:=&-\sum_{a \neq b} \sum_2 \mathcal{K} \int_> \frac{d^2 k_1}{(2 \pi)^2} f_B (\xi_{1a}) (k_{1}^b)^2 \phi_{2 b}^{\ast} \phi_{2 b}\,.\label{part_3}
		\end{align}

		The fast momentum $\mathbf{k}_1=(k_1^1,k_1^2)$ takes values in the elliptic shell ($\Lambda$ is the momentum cut-off and the RG flow parameter $s=e^{l}$ with $s\rightarrow 1$ and $l\rightarrow 0$):  
		\begin{align}
			\left(\frac{\Lambda}{s}\right)^2<(k_1^a)^2+\frac{\mathcal{T}_1}{\mathcal{T}_0}(k_1^b)^2<\Lambda^2\,\label{shell_123}
		\end{align}
		which defines the domain of integral $\int_>$.   Here we should focus on the integrated domain of fast momenta. It is only determined by the fields carrying those fast momenta. To be specific, we consider the fast momentum $\mathbf{k_i}$ carried by the fast mode $\phi_{ia}$ or $\phi_{ia}^{*}$ with the label $i=1,2,...,4$ defined above. Its integrated domain is given by $(\Lambda/s)^2<(k_i^a)^2+\frac{\mathcal{T}_1}{\mathcal{T}_0}(k_i^b)^2<\Lambda^2$ for $b\neq a$. This conclusion will be used in the following text.  To compute  $\langle S_{{int}} \rangle_{>}^{(1)}$ in Eq.~(\ref{part_1}), we may introduce a new momentum $\tilde{\mathbf{k}}$:
		$\tilde{k}^a={k}_1^a\,,\tilde{k}^b=\sqrt{\frac{\mathcal{T}_1}{\mathcal{T}_0}}k_1^b\,. 
		$ Then, Eq.~(\ref{shell_123}) is changed to:
		$\left(\frac{\Lambda}{s}\right)^2<(\tilde{k}^a)^2+(\tilde{k}^b)^2<\Lambda^2\,,\text{ i.e., }\, \left(\frac{\Lambda}{s}\right)^2<\tilde{\mathbf{k}}^2<\Lambda^2\,.
		$ 
		And Eq.~(\ref{eq_dispersion}) is changed to:
		\begin{align}
			\xi_{1a} =
			\frac{\mathcal{T}_0 \left[(\tilde{k}^a)^2 + (\tilde{k}^b)^2\right]}{2}-\mu=\frac{\mathcal{T}_0 \tilde{\mathbf{k}}^2}{2}-\mu\,. \label{eq_dispersion_2}
		\end{align}
		Using the new momentum variables, we can work out the integral $\int_>$ in  
		$\langle S_{{int}} \rangle_{>}^{(1)}$ in Eq.~(\ref{part_1}) ($s-1$ is small enough):
		\begin{align}
			&\int_> \frac{d^2 k_1}{(2 \pi)^2} f_B (\xi_{1a})
			\nonumber\\
			=&\sqrt{\frac{\mathcal{T}_0}{\mathcal{T}_1}} \int_> \frac{d^2 \tilde{k}}{(2 \pi)^2} f_B (\xi_{1a})\nonumber\\
			=&\sqrt{\frac{\mathcal{T}_0}{\mathcal{T}_1}} \frac{1}{(2\pi)^2} \int^{\Lambda}_{\Lambda/s} 2\pi |\tilde{\mathbf{k}}| d |\tilde{\mathbf{k}}|  \frac{1}{e^{\beta\xi_{1a}}-1}  \nonumber\\
			=&\sqrt{\frac{\mathcal{T}_0}{\mathcal{T}_1}} \frac{1}{(2\pi)^2} \int^{\Lambda^2}_{(\Lambda/s)^2} \frac{2\pi}{2}  d (|\tilde{\mathbf{k}}|^2)  \frac{1}{e^{\beta\xi_{1a}}-1} \nonumber\\
			\approx&\sqrt{\frac{\mathcal{T}_0}{\mathcal{T}_1}} \frac{1}{(2\pi)^2}  \frac{2\pi}{2}      \frac{1}{e^{\beta\xi_{0\Lambda}}-1}\left[\Lambda^2-\left(\frac{\Lambda}{s}\right)^2\right]  \nonumber\\
			\approx&\sqrt{\frac{\mathcal{T}_0}{\mathcal{T}_1}} \frac{1}{(2\pi)^2}  \frac{2\pi}{2}      \frac{1}{e^{\beta\xi_{0\Lambda}}-1} 2\Lambda^2 l =\sqrt{\frac{\mathcal{T}_0}{\mathcal{T}_1}} \frac{1}{2\pi}   \frac{1}{e^{\beta\xi_{0\Lambda}}-1} \Lambda^2 l   \,,\nonumber
		\end{align}
		where we have considered infinitesimal $s-1$ and the definition of $s=e^l$. And, $\xi_{0\Lambda}:=\frac{\mathcal{T}_0}{2}\Lambda^2-\mu$. As a result,
		\begin{align}
			\!	\!	\!	\!\!	\!	\langle S_{{int}} \rangle_{>}^{(1)}&= -S_2 f_B (\xi_{0\Lambda}) \frac{\Lambda^2}{(2 \pi)^2}\sum_{a \neq b} \sum_2 \mathcal{K} (k_2^a)^2 \phi_{2 b}^{\ast} \phi_{2 b} l\,,
			\label{A3} 
		\end{align}
		where $S_2:=2\pi\sqrt{\frac{\mathcal{T}_0}{\mathcal{T}_1}}$ and $f_B(\xi_{o\Lambda}):= \frac{1}{e^{\beta\xi_{0\Lambda}}-1}$. 
		Below, we will use $dl$  to replace $l$ since it is infinitesimal.

		The second one $\bra S_{int}\ket_{>}^{(2)}$  in Eq.~(\ref{part_2}) vanishes since it is an odd function of the integrated momentum $\mathbf{k}_1$. The third term in Eq.~(\ref{part_3})  is directly connected to correction of the chemical potential. We just briefly give our results here for it is not our focus:	
		\begin{align}
			\langle S_{{int}} \rangle^{(3)}_{>} & = -\frac{1}{2} \sum_{a \neq b} \sum_2 \Lambda^{4} \frac{S_{2}}{(2\pi)^2} f_B (\xi_{a \Lambda}) K \phi_{2 b}^{\ast} \phi_{2 b} \left( 1- \frac{1}{s} \right)\,. \nonumber
		\end{align}
		
		Therefore, the only term that can renormalize $\mathcal{T}_1$ is $\langle S_{{int}} \rangle^{(1)}_{>} $ in Eq.~(\ref{A3}).   The bare off-diagonal kinetic term is given by 
		$		\sum_{a\neq b} \sum_k\frac{\mathcal{T}_1}{2}(k_2^a)^2\phi_{2 b}^{\ast} \phi_{2 b}\,
		$
		in the frequency-momentum space.  Hence, we have the $\beta$-function of the parameter $\mathcal{T}_1$ by referencing the effective action: $S_{eff}[\phi_<] =S_0[\phi_<]+\bra S_{int}\ket_>[\phi_<]-\frac{1}{2}(\bra S_{int}^2\ket_>-\bra S_{int}\ket_>^2)[\phi_<]$.
		\begin{align}
			\!\!\!\!	\frac{d \mathcal{T}_{1}}{d l} & = - S_2 \mathcal{K} f_B (\xi_{a \Lambda})\frac{\Lambda^2}{(2 \pi)^2}  = - \sqrt{\frac{\mathcal{T}_0}{\mathcal{T}_1}} \mathcal{K} f_B (\xi_{a \Lambda})\frac{\Lambda^2}{2 \pi}\,.\label{A4}
		\end{align}	
		This corresponds to   Fig.~\ref{fig2}. Besides, by referencing the term containing the chemical potential, we have the $\beta$-function of the chemical potential $\mu$: $\frac{d\mu}{dl} = 2\mu - \frac{1}{2}\Lambda^{4} \frac{S_{2}}{(2\pi)^2} f_B (\xi_{a \Lambda}) \mathcal{K}
		$ with the term $2\mu$ originating from the contribution from the tree-level diagrams.

		\subsection{Vertex correction to $\mathcal{K} $}	
		
		We then turn to the vertex correction with all the contractions contained in $\langle S^2_{{int}} \rangle_{>}$: 
		\begin{align}
			\langle S^2_{{int}} \rangle_{>} = & \frac{1}{4\beta^2} \sum_{a \neq b}
			\sum_{a' \neq b'}\, \sum_{1,2,3,4}\,\,\sum_{1',2',3',4'}\nonumber\\
			& K_{{ab}} K_{a' b'}
			\langle \phi_{1 a}^{\ast} \phi_{2 b}^{\ast}
			\phi_{3 b} \phi_{4 a} \phi_{1' a'}^{\ast} \phi_{2' b'}^{\ast} \phi_{3' b'}
			\phi_{4' a'}\rangle_{>} \nonumber\\
			& \cdot (k_2^a + k_1^b) (k_3^a + k_4^b)(k_{2'}^{a'} + k_{1'}^{b'}) (k_{3'}^{a'} + k_{4'}^{b'})\nonumber\\
			&\cdot \delta (1 + 2 -
			3 - 4) \delta (1' + 2' - 3' - 4')\nonumber\\
			=& \langle S^2_{{int}} \rangle_{>}^{(1)}+\langle S^2_{{int}} \rangle_{>}^{(2)}\label{A23} 
		\end{align}
		Since we only consider the loop-diagram contribution, only two kinds of contractions will not vanish under calculation. The first term $\langle S^2_{{int}} \rangle_{>}^{(1)}$ can be concluded as four cases:
		\begin{itemize}
			\item \textbf{Case 1}: The momenta $\mathbf{k_1}$, $\mathbf{k_2}$, $\mathbf{k_{3'}}$ and $\mathbf{k_{4'}}$ are fast momenta. $\mathbf{k_1}=\mathbf{k_{3'}}$, $\omega_{n_1}=\omega_{n_{3'}}$; $\mathbf{k_2}=\mathbf{k_{4'}}$ and  $\omega_{n_2}=\omega_{n_{4'}}$.
			\item \textbf{Case 2}: The momenta $\mathbf{k_1}$, $\mathbf{k_2}$, $\mathbf{k_{3'}}$ and $\mathbf{k_{4'}}$ are fast momenta. $\mathbf{k_1}=\mathbf{k_{4'}}$, $\omega_{n_1}=\omega_{n_{4'}}$; $\mathbf{k_2}=\mathbf{k_{3'}}$ and  $\omega_{n_2}=\omega_{n_{3'}}$.
			\item \textbf{Case 3}: The momenta $\mathbf{k_{1'}}$, $\mathbf{k_{2'}}$, $\mathbf{k_3}$ and $\mathbf{k_4}$ are fast momenta. $\mathbf{k_{1'}}=\mathbf{k_3}$, $\omega_{n_{1'}}=\omega_{n_3}$; $\mathbf{k_{2'}}=\mathbf{k_4}$ and  $\omega_{n_{2'}}=\omega_{n_4}$.
			\item \textbf{Case 4}: The momenta $\mathbf{k_{1'}}$, $\mathbf{k_{2'}}$, $\mathbf{k_3}$ and $\mathbf{k_4}$ are fast momenta. $\mathbf{k_{1'}}=\mathbf{k_4}$, $\omega_{n_{1'}}=\omega_{n_4}$; $\mathbf{k_{2'}}=\mathbf{k_3}$ and  $\omega_{n_{2'}}=\omega_{n_3}$.
		\end{itemize}
		\begin{figure}[t]
			\centering
			\includegraphics[width=0.45\columnwidth]{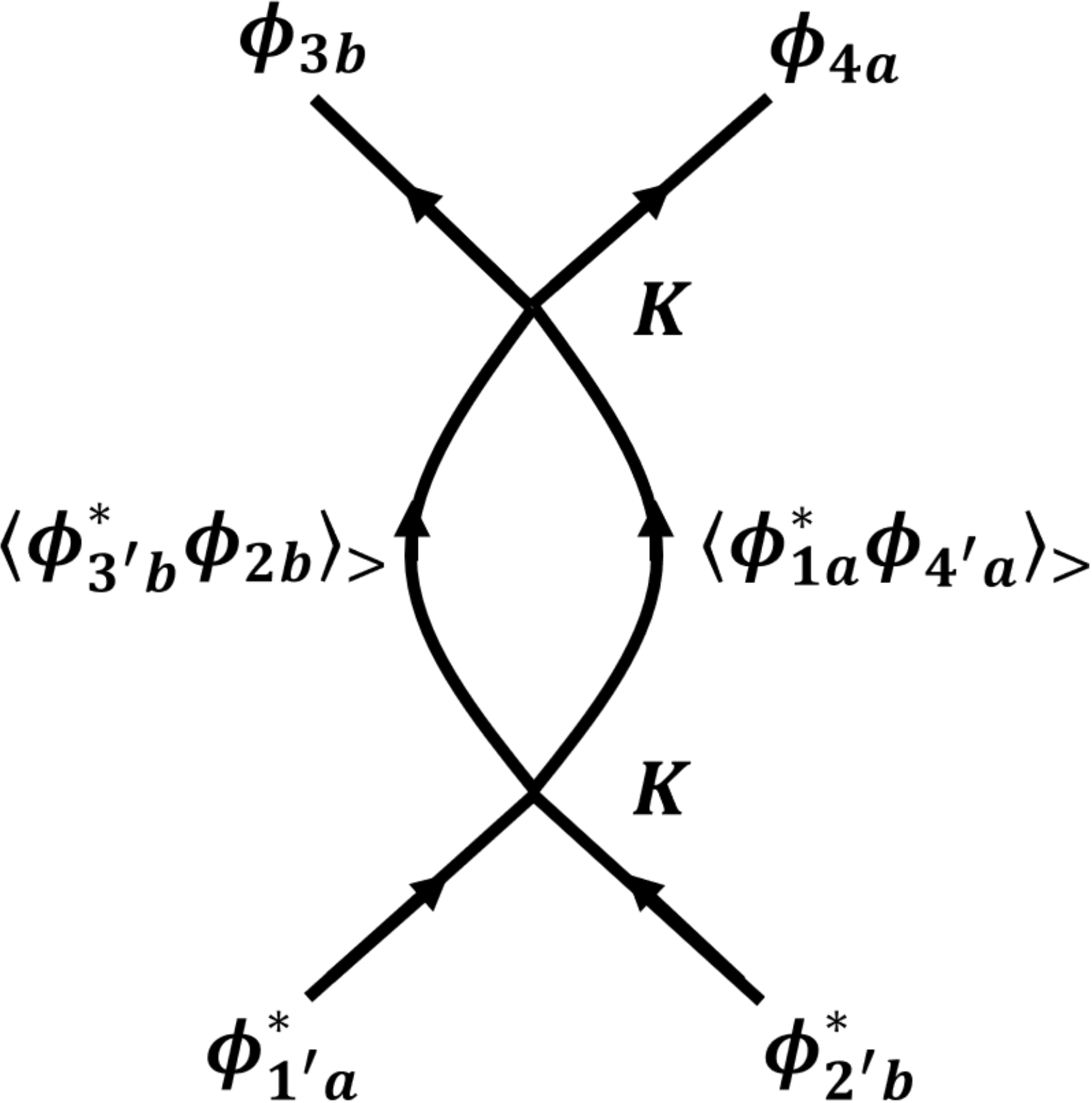}\\
			\caption{The Feynman diagram of the first type of contraction in the vertex correction contributes to the correction of the parameter $\mathcal{K}$. It corresponds to the $\textbf{Case 1}$-$\textbf{Case 4}$ below. For simplicity, we here show one of the related Feynman diagrams. By exchanging the symmetric indices, we can obtain the other three possible diagrams.}
			\label{fig3}
		\end{figure}
		\begin{figure}[t]
			\centering
			\includegraphics[width=0.45\columnwidth]{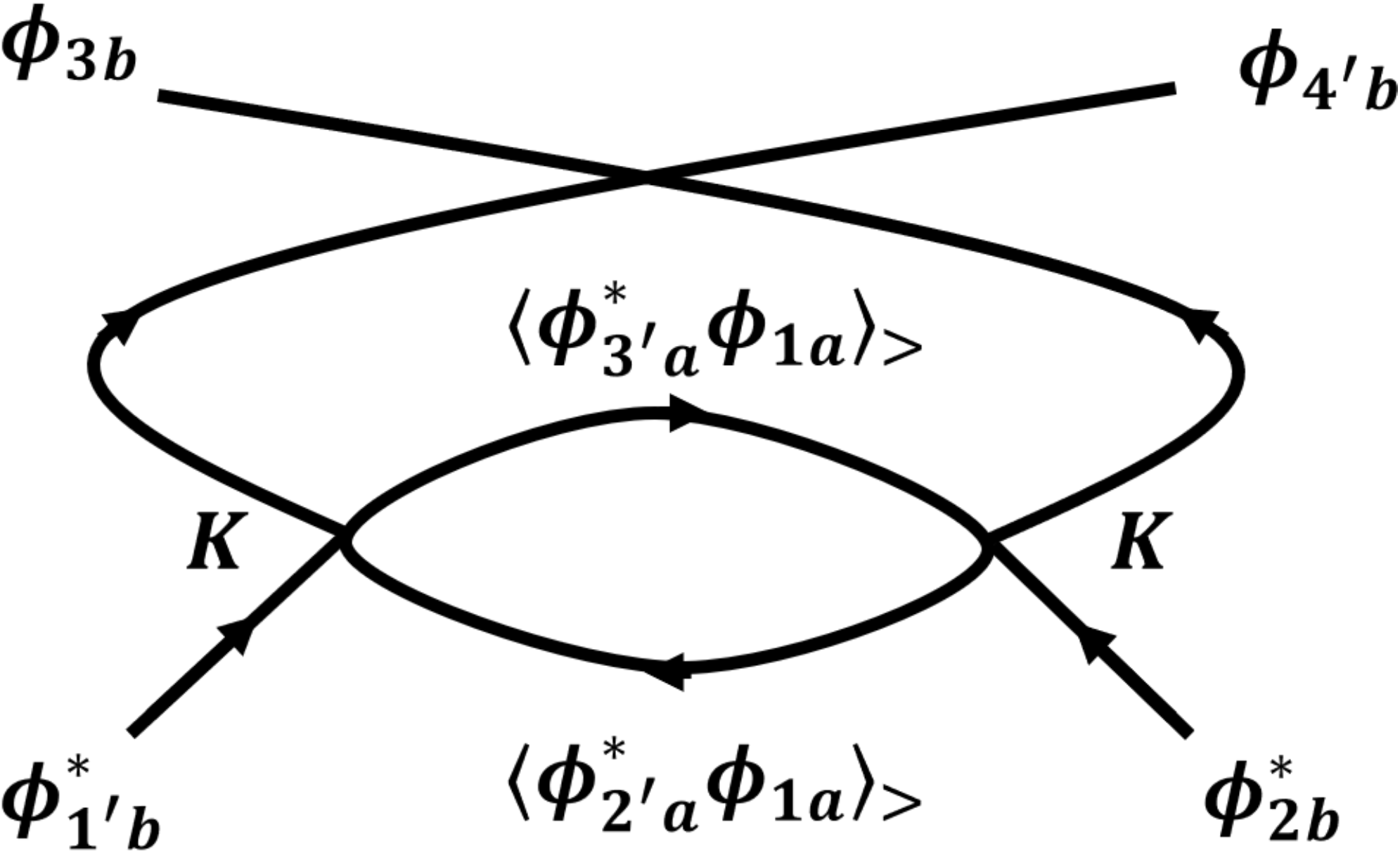}
			\includegraphics[width=0.5\columnwidth]{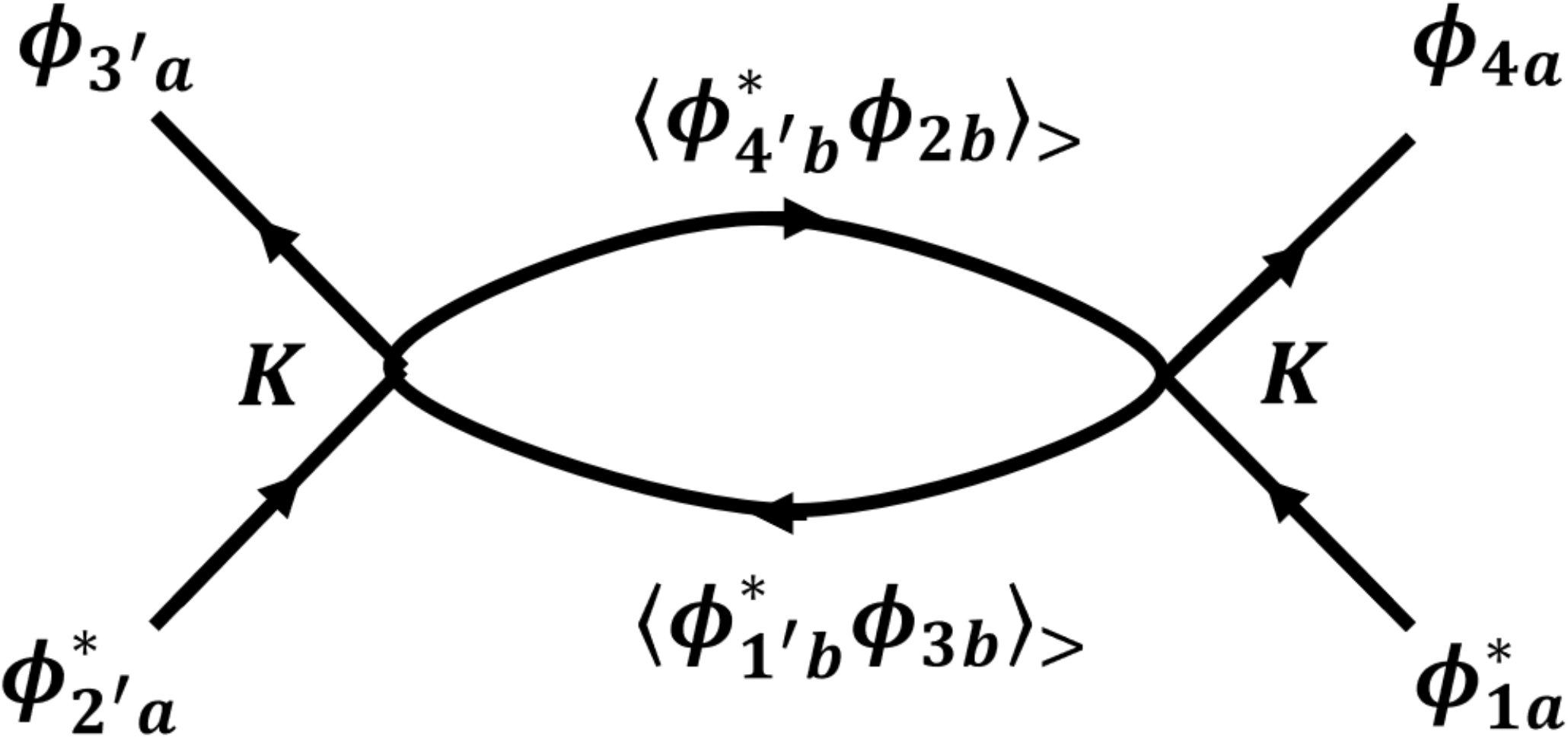}\\
			\caption{The Feynman diagrams of the second type of contraction in the vertex correction contribute to the correction of other terms like $g|\phi_a|^4$. They correspond to the $\textbf{Case 5}$-$\textbf{Case 12}$ below. For simplicity, we here show two of them. By exchanging the symmetric indices, we can have the other six possible diagrams.}
			\label{fig4}
		\end{figure}
		These four cases correspond to the Feynman diagram in Fig. {\ref{fig3}}. Similarly, it can also be proved that these four cases are equal to one another by exchanging the indices. We directly take the $\textbf{Case 1}$ as an example and multiply it by 4 for simplicity. By further considering the delta functions in vertex correction, the relationships between momenta and frequencies are given by: $		\mathbf{k}_1 =\mathbf{k}_{3'}, \mathbf{k}_2 =\mathbf{k}_{4'},
		\mathbf{k}_1 +\mathbf{k}_2 =\mathbf{k}_3 +\mathbf{k}_4
		=\mathbf{k}_{1'} +\mathbf{k}_{2'} =\mathbf{k}_{3'} +\mathbf{k}_{4'}\,;\nonumber\\
		\omega_{n_1} = \omega_{n_{3'}}, \omega_{n_2} = \omega_{n_{4'}}, \omega_{n_1}
		+ \omega_{n_2} = \omega_{n_3} + \omega_{n_4} = \omega_{n_{1'}} +
		\omega_{n_{2'}} = \omega_{n_{3'}} + \omega_{n_{4'}}\,.$
		Hence, with the formula (\ref{A23}), we obtain that
		\begin{widetext}
			\begin{align}
				\langle S_{{int}}^2 \rangle_{>}^{(1)} = & \frac{1}{\beta^2} \sum_{a \neq b}\,
				\sum_{1,2,3,4}\,\, \sum_{1',2'} \mathcal{K}^2 (k_2^a + k_1^b)^2 (k_3^a +
				k_4^b) (k_{2'}^{a'} + k_{1'}^{b'}) \frac{1}{i \omega_{n_1} - \xi_{1 a}}
				\frac{1}{i \omega_{n_2} - \xi_{2 b}} \phi_{1' a}^{\ast} \phi_{2'
					b}^{\ast} \phi_{3 b} \phi_{4 a} \times \nonumber\\
				& \delta (1' + 2' - 3 - 4)\delta(1'+2'-2-1) \nonumber\\
				= & \frac{1}{\beta^2} \sum_{a \neq b} \,\sum_{2,3,4} \,\,\sum_{1', 2'} \mathcal{K}^2
				(k_2^a - k_2^b + k_{1'}^b + k_{2'}^b)^2 (k_3^a + k_4^b) (k_{2'}^{a'} +
				k_{1'}^{b'}) \times \nonumber\\
				& \frac{1}{i \omega_{n_{1'}} + i \omega_{n_{2'}} - i \omega_{n_2} -
					\xi_{1' + 2' - 2, a}} \cdot\frac{1}{i \omega_{n_2} - \xi_{2b}} \phi_{1'
					a}^{\ast} \phi_{2' b}^{\ast} \phi_{3 b} \phi_{4 a} \delta (1' + 2' - 3 - 4)
				\nonumber\\
				= & -\frac{1}{\beta^2} \sum_{a \neq b} \int \frac{d^2 k_2}{(2 \pi)^2}\,\sum_{3,4} \,\,\sum_{1', 2'} \mathcal{K}^2
				(k_2^a - k_2^b + k_{1'}^b + k_{2'}^b)^2 (k_3^a + k_4^b) (k_{2'}^{a'} +
				k_{1'}^{b'})\nonumber\\
				& \sum_{n_2} \frac{1}{i \omega_{n_2} - i \omega_{n_{1'}} - i \omega_{n_{2'}} +
					\xi_{1' + 2' - 2, a}} \cdot\frac{1}{i \omega_{n_2} - \xi_{2b}} \phi_{1'
					a}^{\ast} \phi_{2' b}^{\ast} \phi_{3 b} \phi_{4 a} \delta (1' + 2' - 3 - 4)\,,
				\label{31}
		\end{align}\end{widetext}
		where we introduce the notation: $\xi_{1' + 2' - 2, a}=\frac{\sum_{a'=1}^2t_{aa'}(k_{1'}^{a'}+k_{2'}^{a'}-k_2^{a'})^2}{2}-\mu$. This expression contains a sum of Matsubara frequencies $n_2$: $\sum_{n_2} \frac{1}{i \omega_{n_2} - i \omega_{n_{1'}} - i \omega_{n_{2'}} +
			\xi_{1' + 2' - 2, a}} \cdot\frac{1}{i \omega_{n_2} - \xi_{2b}}$. By applying the formula:
		$	\frac{1}{\beta}\sum_n\frac{1}{i\omega_{n}-\xi_1}\frac{1}{i\omega_{n}-\xi_2}=-\frac{f_B(\xi_1)-f_B(\xi_2)}{\xi_1-\xi_2}  $
		for Matsubara frequencies for bosons: $\omega_n =\frac{2 \pi n}{\beta}$, we can obtain the result with substitutions: $\xi_1=i\omega_{n_{1'}}+i\omega_{n_{2'}}-\xi_{1'+2'-2, a}$ and $\xi_2=\xi_{2b}$. The sum of $n_2$ can be rewritten as:
		\begin{align}
			&\frac{1}{\beta}\sum_{n_2} \frac{1}{i \omega_{n_2} - i \omega_{n_{1'}} - i \omega_{n_{2'}} +
				\xi_{1' + 2' - 2, a}} \cdot\frac{1}{i \omega_{n_2} - \xi_{2b}} \nonumber\\
			=	&	 -\frac{f_B (\xi_{2 b}) - f_B (-\xi_{1'+2'-2,a})}{\xi_{1'+2'-2,a} + \xi_{2 b} - i \omega_{n_{1'}} - i \omega_{n_{2'}}}\nonumber\\
			=	&  -\frac{f_B (\xi_{2 b}) + f_B (\xi_{1'+2'-2,a})+1}{\xi_{1'+2'-2,a} + \xi_{2 b} - i \omega_{n_{1'}} - i \omega_{n_{2'}}}\,,
			\label{33}
		\end{align}
		where we apply 
		$f_B (i \omega_{n_1} + i \omega_{n_2} - \xi) 
		= \frac{1}{\exp [2 \pi i (n_1 +
			n_2) - \beta \xi] - 1} = \frac{1}{\exp [-\beta \xi] - 1}
		= f_B(-\xi)=\frac{1}{e^{-\beta\xi}-1} =-1-\frac{1}{e^{\beta\xi}-1}=-1-f_B(\xi)\,.
		$ 
		Replacing the sum in (\ref{31}) with (\ref{33}), $\langle S_{{int}}^2 \rangle_{>}^{(1)}$ can be rewritten as:
		\begin{align}
			\langle S_{{int}}^2 \rangle_{>}^{(1)} = & \frac{1}{\beta}\sum_{a \neq b} \sum_{3,4} \sum_{1', 2'} 
			\mathcal{K}^2(k_3^a + k_4^b) (k_{2'}^{a} + k_{1'}^{b}) \nonumber\\
			&\cdot\phi_{1'a}^{\ast} \phi_{2'b}^{\ast} \phi_{3 b} \phi_{4 a}\delta (1'+2'-3-4)  
			\nonumber\\
			&\cdot \int \frac{d^2 k_2}{(2 \pi)^2} \frac{f_B (\xi_{2 b}) + f_B (\xi_{1'+2'-2,a})+1}{\xi_{1'+2'-2,a} + \xi_{2 b} - i \omega_{n_{1'}} - i \omega_{n_{2'}}}\nonumber\\
			&\cdot  (k_2^a - k_2^b + k_{1'}^b + k_{2'}^b)^2\nonumber\\
			\approx & \frac{1}{\beta}\sum_{a \neq b} \sum_{3,4} \sum_{1', 2'} \mathcal{K}^2(k_3^a + k_4^b) (k_{2'}^{a'} + k_{1'}^{b'})\nonumber\\
			&\cdot \phi_{1' a}^{\ast} \phi_{2'b}^{\ast} \phi_{3 b} \phi_{4 a} \delta (1' + 2' - 3 - 4)  \nonumber\\
			& \int \frac{d^2 k_2}{(2 \pi)^2} \frac{f_B (\xi_{2 b}) + f_B (\xi_{2a})+1}{\xi_{2 a} + \xi_{2 b} - i \omega_{n_{1'}} - i \omega_{n_{2'}}} (k_2^a - k_2^b)^2\,,\nonumber
		\end{align}
		where the momenta $\mathbf{k_{1'}}$ and $\mathbf{k_{2'}}$ can be ignored in the expression since $\phi_{1'a}$ and $\phi_{2'b}$ are both slow modes. 
		The integrated region of the momentum $\mathbf{k_2}$ in $\langle S_{{int}}^2 \rangle_{>}^{(1)}$ is given by: $(\Lambda/s)^2\leq\sum_{a(a\neq b)}(k_2^b)^2+\frac{\mathcal{T}_1}{\mathcal{T}_0}(k_2^a)^2\leq\Lambda^2$ since the momentum $\mathbf{k_2}$ is carried by the field $\phi_{2b}^{*}$. With substitutions $\tilde{k}^b=k_2^b,\tilde{k}^a=\sqrt{\frac{\mathcal{T}_1}{\mathcal{T}_0}}k_2^a$, we limit our integrated region on $(\Lambda/s)^2\leq\tilde{\mathbf{k}}^2\leq\Lambda^2$. Therefore, the contribution from the first type of contraction is given by:
		\begin{align}
			\langle S_{{int}}^2 \rangle_{>}^{(1)} =& \frac{A}{\beta}\sum_{a \neq b} \sum_{1'2'34}\mathcal{K}^2(k_3^a + k_4^b) (k_{2'}^{a} + k_{1'}^{b}) \nonumber\\
			&\cdot\phi_{1' a}^{\ast} \phi_{2'b}^{\ast} \phi_{3 b} \phi_{4 a}\delta (1' + 2' - 3 - 4)dl\,,\label{A25}
		\end{align}
		where the parameter $A$ takes the form of: 
		
		\begin{align}
			\!\!		A &=\frac{1}{dl} \int_{>} \frac{d^{2} k_2}{(2 \pi)^2} \frac{ f_B(\xi_{2a})+f_B (\xi_{2b})+1 }{\xi_{2a} + \xi_{2b} - \omega_{1'} -\omega_{2'}-i0^{+}} (k_{2}^a-k_{2}^b)^2\nonumber\\
			&\approx\frac{1}{dl} \int_{>} \frac{d^{2} k_2}{(2 \pi)^2} \frac{ f_B(\xi_{2a})+f_B (\xi_{2b})+1 }{\xi_{2a} + \xi_{2b}} (k_{2}^a-k_{2}^b)^2\,,\label{A}
		\end{align}
		where we use analytic continuation here and replace the imaginary frequencies with $\omega_{1'}$ and $\omega_{2'}$. It is necessarily assume that the frequencies and momenta on the external lines can be ignored compared with $\xi_{0\Lambda}$ and $\Lambda$. It is obvious that the expression (\ref{A}) is positive. In other words, $A>0$. More calculations on $A$ are shown in Appendix~\ref{appendix_parameter_a}. By comparing the form of (\ref{A25}) with the bare interaction vertex,		we find that it will correct the interaction parameter $\mathcal{K}$ effectively. According to the effective action:$S_{eff}=S_0+\bra S_{int}\ket-\frac{1}{2}(\bra S_{int}^2\ket-\bra S_{int}\ket^2)$, we have the form of the $\beta$-function of parameter $\mathcal{K}$:
		\begin{align}
			\frac{d \mathcal{K}}{dl} = -\mathcal{K}^2 \times A- 2\mathcal{K}\,,
			\label{A14}
		\end{align}
		where the term $-2\mathcal{K}$ originates from the rescaling of slow momenta carried by the slow modes in (\ref{K}) above.  We have the general form of $\mathcal{K}(l)$ and $\mathcal{T}_1(l)$:
		\begin{align}
			\mathcal{K}(l) =& \frac{2/A}{\frac{\mathcal{K}(0)+2/A}{\mathcal{K}(0)}e^{2l}-1}\,,\\
			\mathcal{T}_{1}(l) =& \bigg[\mathcal{T}_{1}(0)^{3/2}+\sqrt{\mathcal{T}_0}\frac{3\Lambda^2}{4A\pi} f_{B} (\xi_{0\Lambda})\nonumber\\
			&\cdot\ln(\frac{2}{-\mathcal{K}(0)Ae^{-2l}+(2+\mathcal{K}(0)A)})\bigg]^{2/3}\,,
		\end{align}
		 with the separatrix:
			\begin{equation}
				\mathcal{T}_{1}(0)=(f_{B} (\xi_{0\Lambda})\sqrt{\mathcal{T}_0}\frac{3\Lambda^2}{4A\pi}\ln(\frac{2+\mathcal{K}(0)A}{2}))^{2/3}\,.\label{separatrix}
		\end{equation}

		The second type of contraction $\langle S_{{int}}^2 \rangle_{>}^{(2)}$ contains overall eight cases presented below.
		\begin{itemize}
			\item \textbf{Case 5}: The momenta $\mathbf{k_1}$, $\mathbf{k_3}$, $\mathbf{k_{2'}}$ and $\mathbf{k_{4'}}$ are fast momenta. $\mathbf{k_1}=\mathbf{k_{4'}}$, $\omega_{n_1}=\omega_{n_{4'}}$; $\mathbf{k_3}=\mathbf{k_{2'}}$ and  $\omega_{n_3}=\omega_{n_{2'}}$.
			\item \textbf{Case 6}: The momenta $\mathbf{k_1}$, $\mathbf{k_4}$, $\mathbf{k_{1'}}$ and $\mathbf{k_{4'}}$ are fast momenta. $\mathbf{k_1}=\mathbf{k_{4'}}$, $\omega_{n_1}=\omega_{n_{4'}}$; $\mathbf{k_4}=\mathbf{k_{1'}}$ and  $\omega_{n_4}=\omega_{n_{1'}}$.
			\item \textbf{Case 7}: The momenta $\mathbf{k_1}$, $\mathbf{k_3}$, $\mathbf{k_{1'}}$ and $\mathbf{k_{3'}}$ are fast momenta. $\mathbf{k_1}=\mathbf{k_{3'}}$, $\omega_{n_1}=\omega_{n_{3'}}$; $\mathbf{k_3}=\mathbf{k_{1'}}$ and  $\omega_{n_3}=\omega_{n_{1'}}$.
			\item \textbf{Case 8}: The momenta $\mathbf{k_1}$, $\mathbf{k_4}$, $\mathbf{k_{2'}}$ and $\mathbf{k_{3'}}$ are fast momenta. $\mathbf{k_1}=\mathbf{k_{3'}}$, $\omega_{n_1}=\omega_{n_{3'}}$; $\mathbf{k_4}=\mathbf{k_{2'}}$ and  $\omega_{n_4}=\omega_{n_{2'}}$.
			\item \textbf{Case 9}: The momenta $\mathbf{k_2}$, $\mathbf{k_3}$, $\mathbf{k_{2'}}$ and $\mathbf{k_{3'}}$ are fast momenta. $\mathbf{k_2}=\mathbf{k_{3'}}$, $\omega_{n_2}=\omega_{n_{3'}}$; $\mathbf{k_3}=\mathbf{k_{2'}}$ and  $\omega_{n_3}=\omega_{n_{2'}}$.
			\item \textbf{Case 10}: The momenta $\mathbf{k_2}$, $\mathbf{k_4}$, $\mathbf{k_{1'}}$ and $\mathbf{k_{3'}}$ are fast momenta. $\mathbf{k_2}=\mathbf{k_{3'}}$, $\omega_{n_2}=\omega_{n_{3'}}$; $\mathbf{k_4}=\mathbf{k_{1'}}$ and  $\omega_{n_4}=\omega_{n_{1'}}$.
			\item \textbf{Case 11}: The momenta $\mathbf{k_2}$, $\mathbf{k_3}$, $\mathbf{k_{1'}}$ and $\mathbf{k_{4'}}$ are fast momenta. $\mathbf{k_2}=\mathbf{k_{4'}}$, $\omega_{n_2}=\omega_{n_{4'}}$; $\mathbf{k_3}=\mathbf{k_{1'}}$ and  $\omega_{n_3}=\omega_{n_{1'}}$.
			\item \textbf{Case 12}: The momenta $\mathbf{k_2}$, $\mathbf{k_4}$, $\mathbf{k_{2'}}$ and $\mathbf{k_{3'}}$ are fast momenta. $\mathbf{k_2}=\mathbf{k_{3'}}$, $\omega_{n_2}=\omega_{n_{3'}}$; $\mathbf{k_4}=\mathbf{k_{2'}}$ and  $\omega_{n_4}=\omega_{n_{2'}}$.
		\end{itemize}
		
		The Feynman diagrams related are presented in Fig. {\ref{fig4}}. After integrating the fast momenta, all of these terms will be turned into terms with constant coefficients. Therefore, integration from the two connected diagrams will be explained as corrections of other possible terms such as $g | \phi_a |^4$. We do not care about this parameter for it does no contribution to the parameters $\mathcal{K}$ and $\mathcal{T}_{1}$. For simplicity, we do not present more details here.

		\subsection{Global phase diagram}

		In summary,  we can see the first-order corrections only correct the kinetic terms on directions other than the $a$-th one of field configurations $\phi_a$  [see definition below Eq.~(\ref{equation_action_Main})].  In this way, the kinetic energy along the $a$-th axis will not be influenced by the contraction. We can directly prove that the higher-order correction will still contribute nothing to the parameter $\mathcal{T}_0$. Hence, we have the $\beta$-function for parameter $\mathcal{T}_{1}$  corresponding to Fig. {\ref{fig2}}, given by Eq.~(\ref{A4}). 
		We can safely come to the conclusion that if $\mathcal{K} > 0$, the
		parameter $t_{1}$ will reduce in the RG flow. This requires the parameter $\mathcal{K}$ to be positive. Only with positive $\mathcal{K}$ can parameter $\mathcal{T}_1$ flow to zero in the RG analysis. Nevertheless, at this step, it is insufficient to tell whether the parameter $\mathcal{T}_1$ is irrelevant here since the elements in the matrix $\mathcal{K}$ will also flow to zero. What we need is further calculation on vertex correction, which has been given in  Eq.~(\ref{A14}).

		\begin{figure}[t]
			\centering
			\includegraphics[width=0.95\columnwidth]{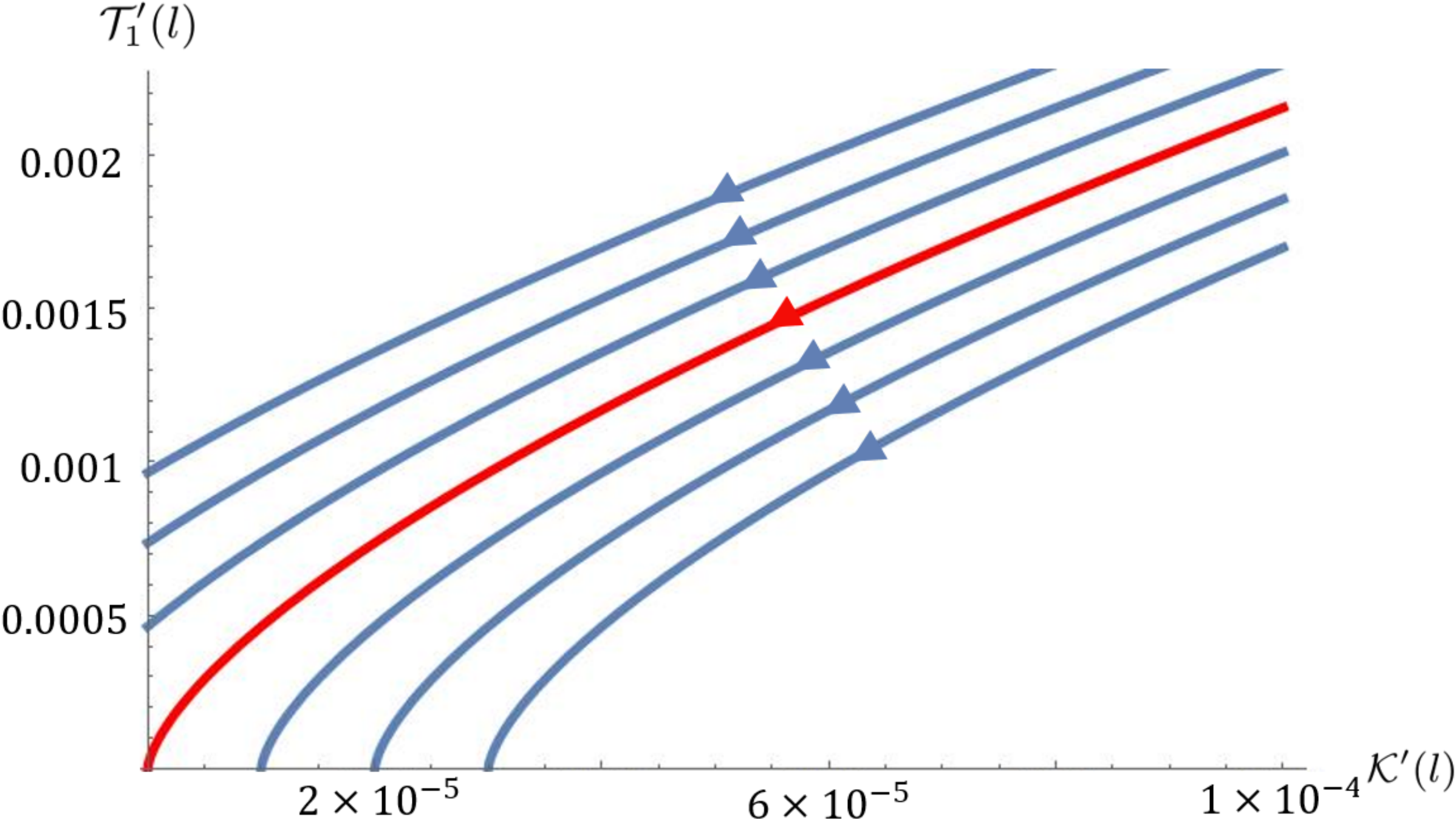}\\
			\caption{RG flow for $\mathcal{T}_{1}'(l)= {\mathcal{T}_{1}(l)}{\left(f_B(\xi_{0\Lambda})\sqrt{\mathcal{T}_0}\frac{3\Lambda^2}{4A\pi}\right)^{-2/3}}$ and $\mathcal{K}'(l)= {A\mathcal{K}(l)}/{2}$. Here the red line is a separatrix below which higher-rank symmetry emerges. The parameters $(\mathcal{T}_1'(0),\mathcal{K}'(0))$ of these lines are (from above to below): $(2.57\times10^{-3},10^{-4})$, $(2.43\times10^{-3},10^{-4})$, $(2.30\times10^{-3},10^{-4})$, $(2.16\times10^{-3},10^{-4})$, $(2.01\times10^{-3},10^{-4})$, $(1.86\times10^{-3},10^{-4})$, and $(1.70\times10^{-3},10^{-4})$. }
			\label{figure_RGflow}
		\end{figure}

		A numerical result of beta functions (\ref{A4}) and (\ref{A14}) is shown in Fig.~\ref{figure_RGflow},  where the red line is the separatrix (\ref{separatrix}). In the region below the line, all $\mathcal{T}_{1}(0)$ will flow to zero in RG flow, indicating the emergence of higher-rank symmetry in low-energy physics since those terms violating the higher-rank symmetry will vanish in the low-energy physics. { In addition, after comparing the parameters $\mathcal{T}_1(0)$ and $\mathcal{K}(0)\Lambda^2$ near the separatrix, we have $\mathcal{T}_1(0)\ll \mathcal{K}(0)\Lambda^2\ll \mathcal{T}_0$ with the approximation $\frac{\xi_{0\Lambda}}{2k_BT}\gg1$, which can be realized in   cold-atom systems. This tells us our calculation is consistent with our initial assumption above.}
		
		From the phase diagram, we can safely come to the conclusion that the higher-rank symmetry will emerge after integrating the high-energy modes since off-diagonal kinetic parameters are just perturbations to the system. However, it should be noted that our RG analysis is only valid near the initial point $l=0$. We here also do not take higher-order terms into account. Therefore, our calculation and results may be invalidated when $l\rightarrow\infty$. The reason we do not just finish our RG analysis on the axis: $\mathcal{T}_1(0)=0$ is we only consider the lower-order corrections. If we start with an initial point on the horizontal axis: $\mathcal{T}_1(0)=0$, we have $S_2\rightarrow\infty$ in $\frac{d\mathcal{T}_1}{dl}$, which requires more   higher-order terms since they contain expressions with higher order of $S_2$. We have to take all of them into account for accuracy, which is beyond the current perturbation calculations. {Although our calculation cannot take all the higher-order corrections into account, it also exhibits the tendency of RG flow giving the accurate prediction near the region with $l=0$.} Alternatively, the exact flat ``band'' along one direction for a boson, in analogy to bosonic Landau level, leads to strong correlation effect even though interaction is not large. 
		We also comment that the perturbation calculations fail completely if we instead study  models \cite{FS1} with conserved dipole moments, where bosons are fractons that are immobile towards all directions. In such models, the Hamiltonian is intrinsically non-Gaussian with no kinetic terms at all. But analytic difficulty becomes much smaller when fracton condensation is considered as what Ref.~\cite{FS1} did. 
		
		Above discussion is all about the $\beta$-functions of parameters of $2$-component bosons in two-dimensional space. Similarly, we can extend the case to the $d$-dimensional space with $d>2$. We here give the forms of $\beta$-functions of $d$-component bosonic systems for reference. The $\beta$-functions of ${t}_{ab}(a\neq b)$ and $K_{ab}(a\neq b)$ is given by:   $\frac{d t_{ab}}{d l} = -S_d K_{ab} f_B (\xi_{0\Lambda})\frac{\Lambda^d}{(2 \pi)^d} ,	\frac{d K_{ab}}{d l} = -A_d  K_{ab}^2- d  K_{ab} $, 
		where $S_d$ represents the volume of $d$-dimensional ellipsoid:
		$\sum_{b=1}^d \frac{t_{ab}}{t_a}(k^b)^2=1$  and 
		$A_d\!= \!\frac{1}{2\beta l}\!\!\int_{P_d} \!\!\frac{d^d k}{(2 \pi)^d}\! \frac{f_B(\xi_{ka})+f_B (\xi_{kb})+1}{\xi_{k a} + \xi_{kb}}\! (k^{a}-k^{b})^2\!$
		with integrated domain $P_d: (\Lambda e^{-l})^2\leq\sum_{b=1}^d \frac{t_{ab}}{t_a}(k^b)^2\leq\Lambda^2$. They give the same type of RG flow phase diagram as Fig. {\ref{figure_RGflow}}.  
		
		\section{Summary and Outlook}\label{summary}
		
		In this paper, we study how higher-rank symmetry (\ref{eq_symmetry}) and angular moment conservation (\ref{conserved_angular_total}) emerge at low energies through a RG analysis. In other words, we identify them as emergent phenomena rather than strict properties of microscopic models. A phase diagram is given in Fig.~\ref{figure_RGflow}, in which a wide parameter region is found to support emergent phenomena. {Despite the limitation of one-loop perturbation techniques, we argue that emergence occurs in the deep infrared regime. On the other hand,} by regarding higher-rank symmetry as emergent symmetry, our work opens a door to new way of thinking on   realization of such unconventional symmetry and higher moment conservation in more realistic models, e.g., in simple frustrated spin models near symmetric points. Recently, some higher-moment conserving 1D spin systems have been found to support   anomalously slow, subdiffusive late-time transport \cite{PhysRevLett.125.245303,moudgalya2021spectral}. Thus, it is interesting to ask how   \textit{emergent} conservation of angular moments affect the late-time transport. Besides, the Hamiltonian can be reformulated on a square lattice. In an optical lattice of cold-atom experiments,  one may choose lattice constant $l=700nm,\Lambda\sim\frac{1}{l}=1.43\times10^{7}m^{-1},m_0=\frac{1}{t_0}=8.22\times10^{-34}kg,T=10K$. The condition  $\frac{t_0\Lambda^2-2\mu}{2k_BT}\sim10^3$ where the chemical potential is negative indicates the possibility for simulating the 2-component bosons on the optical lattice.  For accuracy and extension, it will be interesting to further consider the correction from higher-order term with more loops in the Feynman diagram. Finally, it will be interesting to study symmetry-protected topological phases (SPTs) with such emergent higher-rank symmetry.

		\section*{Acknowledgements}
		Discussions with Ruizhi Liu and Yixin Xu are acknowledged.  This work of both authors was done in Guangzhou South Campus of Sun Yat-sen University (SYSU) with full financial support from SYSU talent plan, Guangdong Basic and Applied Basic Research Foundation under Grant No.~2020B1515120100, and    National Natural Science Foundation of China   Grant (No.~11847608 $\&$ No.~12074438).
		
		\appendix
		
		\section{On how  the angular moment conservation affects single particle motion}\label{appendix_a}
		
		In the main text, we introduce the following conserved quantities ($a,b=1,2,\cdots,d\,;a<b$) in $d$-dimensional space:   
		\begin{align}
			\hat{\mathcal{Q}}^a&=\int d^dx\hat{\rho}^a\,,\,
			\hat{\mathcal{Q}}^{ab}=\int d^dx(\hat{\rho}^a x^b-\hat{\rho}^b x^a)\,.
		\end{align} 
		For each $a$, the conservation of $\hat{\mathcal{Q}}^a$ requires that all $a$-th component bosons are always in the $d$-dimensional space. The conserved quantities $\hat{\mathcal{Q}}^{ab}$ leads to the mobility restriction:  a single $a$-th component boson is only allowed to move along $a$-th directions. To be more specific, we consider $d=2$ and the following three quantities are conserved:
		\begin{align}
			\hat{\mathcal{Q}}^1=\int d^dx\hat{\rho}^1\,, \hat{\mathcal{Q}}^2=\int d^dx\hat{\rho}^2\,,\hat{\mathcal{Q}}^{12}=\int d^dx(\hat{\rho}^1 x^2-\hat{\rho}^2 x^1)\,.
		\end{align} 
		$\hat{\mathcal{Q}}^1$ and $\hat{\mathcal{Q}}^2$ are the usual conserved charge (particle number) of bosons of the $1$st and $2$nd components, respectively.   The conserved quantity $\hat{\mathcal{Q}}^{12}$ is the total angular moments formed by bosons. Suppose $\hat{\mathcal{Q}}^1=N_1$ and $\hat{\mathcal{Q}}^2=N_2$. We can use Dirac function to express $\hat{\rho}^1$ and $\hat{\rho}^2$ {in the eigenbasis}:
		\begin{align}
			\rho^1(\mathbf{x})=&\sum^{N_1}_{i}\delta(\mathbf{x}-\mathbf{x}_i) \,, \rho^2(\mathbf{x})=\sum^{N_2}_{j}\delta(\mathbf{x}-\mathbf{x}_j)\,.
		\end{align}
		It means that bosons of the $1$st ($2$nd) component are located in $\mathbf{x}_i$ ($\mathbf{x}_j$) with $i=1,2,3,\cdots, N_1$ ($j=1,2,3,\cdots, N_2$). Then, $\mathcal{Q}^{12}$ are reduced to:
		\begin{align}
			\mathcal{Q}^{12}&= \sum^{N_1}_{i}x^2_i-\sum^{N_2}_{j}x^1_j\,.
		\end{align}
		From this expression, one can conclude that, if we move a $1$st component boson, in order to keep $\mathcal{Q}^{12}$ invariant, the boson is only allowed to be movable in the $1$st direction such that its coordinate $x^2$ is unchanged. Of course, we can \textit{collectively} move bosons of both components, such that the change in $\sum^{N_1}_{i}x^2_i$ can be canceled out by the change in $\sum^{N_2}_{j}x^1_j$ such that $\mathcal{Q}^{12}$ is unaltered. This scenario is beyond the single particle movement and gives nontrivial effects when inter-component interaction ($K$-term) is involved. 
		
		\section{Details of the parameter $A$}\label{appendix_parameter_a}
		Here we focus on the specific value of the parameter $A$ by taking some approximations here. By referencing the definition of $\tilde{\mathbf{k}}$ above, we have:
		\begin{align}
			A &=\frac{1}{dl}\int_{>}\frac{d^{2} \tilde{\mathbf{k}}}{(2 \pi)^2}\frac{ f_B(\xi_{2a})+f_B (\xi_{2b})+1 }{\xi_{2a} + \xi_{2b}} (k_{2}^a-k_{2}^b)^2\sqrt{\frac{\mathcal{T}_0}{\mathcal{T}_1}}\nonumber\\
			&=\frac{1}{dl}\int_{>}\frac{|\tilde{\mathbf{k}}|d\tilde{|\mathbf{k}|}d\theta}{(2 \pi)^2}\frac{f_B(\xi_{2a})+f_B (\xi_{2b})+1}{\xi_{2a} + \xi_{2b}} (k_{2}^a-k_{2}^b)^2\sqrt{\frac{\mathcal{T}_0}{\mathcal{T}_1}}\nonumber\\
			&=\frac{1}{dl}\int_{>}\frac{d|\tilde{\mathbf{k}}|^2d\theta}{2(2 \pi)^2}\frac{ f_B(\xi_{2a})+f_B (\xi_{2b})+1 }{\xi_{2a} + \xi_{2b}} (k_{2}^a-k_{2}^b)^2\sqrt{\frac{\mathcal{T}_0}{\mathcal{T}_1}}\,.\nonumber\\
		\end{align}
		\begin{widetext}
			For simplicity, we will approximately use $\mathcal{T}_1(0)$ to replace $\mathcal{T}_1$ below for convenience. Besides, we have $k_2^b=\Lambda\sin\theta$ and $k_2^a=\Lambda\sqrt{\frac{\mathcal{T}_0}{\mathcal{T}_1}}\cos\theta$ in the integrated domain. Hence, the kinetic energy $\xi_{2a}$ and $\xi_{2b}$ can be approximately considered as $\xi_{1\Lambda}=\xi_{0\Lambda}(\frac{\mathcal{T}_0}{\mathcal{T}_1}\cos^2\theta+\frac{\mathcal{T}_1}{\mathcal{T}_0}\sin^2\theta)$ and $\xi_{0\Lambda}$ seperately if we assume the chemical potential is small enough: $|\mu|\ll\frac{t_0\Lambda^2}{2}$. In this way, the parameter $A$ can be rewritten as:
			\begin{align}
				A \approx&\frac{1}{2dl}\int_{0}^{2\pi}\frac{d\theta}{(2 \pi)^2}(\Lambda^2-(\frac{\Lambda}{s})^2)\frac{f_B(\xi_{1\Lambda})+f_B (\xi_{0\Lambda})+1}{\xi_{1\Lambda} + \xi_{0\Lambda}}(\Lambda\sqrt{\frac{\mathcal{T}_0}{\mathcal{T}_1}}\cos\theta-\Lambda \sin\theta)^2\sqrt{\frac{\mathcal{T}_0}{\mathcal{T}_1(0)}}\nonumber\\
				=&\Lambda^4\int_0^{2\pi} \frac{d\theta}{(2 \pi)^2} \frac{f_B(\xi_{1\Lambda})+f_B (\xi_{0\Lambda})+1}{\xi_{1\Lambda} + \xi_{0\Lambda}} (\sqrt{\frac{\mathcal{T}_0}{\mathcal{T}_1}}\cos\theta-\sin\theta)^2\sqrt{\frac{\mathcal{T}_0}{\mathcal{T}_1(0)}}\nonumber\\
				=&\Lambda^4\int_0^{2\pi} \frac{d\theta}{(2 \pi)^2} \frac{f_B(\xi_{1\Lambda})+f_B (\xi_{0\Lambda})+1}{\xi_{1\Lambda} + \xi_{0\Lambda}} (\sqrt{\frac{\mathcal{T}_0}{\mathcal{T}_1}}\cos\theta-\sin\theta)^2\sqrt{\frac{\mathcal{T}_0}{\mathcal{T}_1(0)}}\nonumber\\
				\approx&\Lambda^4\int_0^{2\pi}\frac{d\theta}{(2\pi)^2}\frac{1}{(\xi_{0\Lambda}+\xi_{1\Lambda})(e^{\xi_{0\Lambda}(\frac{\mathcal{T}_0}{\mathcal{T}_1}\cos^2\theta+\frac{\mathcal{T}_1}{\mathcal{T}_0}\sin^2\theta)/k_BT}-1)}\sqrt{\frac{\mathcal{T}_0}{\mathcal{T}_1(0)}}\nonumber\\
				&+\Lambda^4\sqrt{\frac{\mathcal{T}_0}{\mathcal{T}_1(0)}}\int_0^{2\pi}\frac{d\theta}{(2\pi)^2}\frac{f_B(\xi_{0\Lambda})+1}{\xi_{1\Lambda}+\xi_{0\Lambda}}(\sqrt{\frac{\mathcal{T}_0}{\mathcal{T}_1}}\cos\theta-\sin\theta)^2\nonumber\\
				\approx &\frac{4\Lambda^2}{\mathcal{T}_0}\int_{\frac{\pi}{2}-\mathcal{T}_1^{'}}^{\frac{\pi}{2}+\mathcal{T}_1^{'}}\frac{d\theta}{(2\pi)^2}\frac{2k_BT}{\mathcal{T}_0\Lambda^2\mathcal{T}_1^{'}\sin^2\theta}\sqrt{\frac{\mathcal{T}_0}{\mathcal{T}_1(0)}}+\Lambda^4\sqrt{\frac{\mathcal{T}_0}{\mathcal{T}_1(0)}}\int_0^{2\pi}\frac{d\theta}{(2\pi)^2}\frac{(\sqrt{\frac{\mathcal{T}_0}{\mathcal{T}_1}}\cos\theta-\sin\theta)^2(1+f_B(\xi_{0\Lambda}))}{\xi_{0\Lambda}(1+\frac{\mathcal{T}_0}{\mathcal{T}_1}\cos^2\theta+\frac{\mathcal{T}_1}{\mathcal{T}_0}\sin^2\theta)}\,,
			\end{align}
			
		\end{widetext}
		with $\mathcal{T}_1^{'}=\frac{\mathcal{T}_1}{\mathcal{T}_0}\ll1$. Here we approximately consider that $\frac{\mathcal{T}_1\Lambda^2}{2k_BT}\ll1$. This expression contains two integrals. The first one includes a trick. Since $\mathcal{T}_1^{'}\ll1$, the first integral becomes nontrivial only around the two points: $\theta=\frac{\pi}{2}$ and $\theta=\frac{3\pi}{2}$, which give the same value. Therefore, we change our integrated region as: $\int_0^{2\pi}\rightarrow2\int_{\frac{\pi}{2}-\mathcal{T}_1^{'}}^{\frac{\pi}{2}+\mathcal{T}_1^{'}}$. In addition, we have $\xi_{1\Lambda}=\frac{\mathcal{T}_1\Lambda^2}{2}\ll\xi_{0\Lambda}$ at $\theta=\frac{\pi}{2}$ and $\theta=\frac{3\pi}{2}$. In this way, the first integral given by: $\frac{4\Lambda^2}{\mathcal{T}_0}\int_{\frac{\pi}{2}-\mathcal{T}_1^{'}}^{\frac{\pi}{2}+\mathcal{T}_1^{'}}\frac{d\theta}{(2\pi)^2}\frac{2k_BT}{\mathcal{T}_0\Lambda^2\mathcal{T}_1^{'}\sin^2\theta}\sqrt{\frac{\mathcal{T}_0}{\mathcal{T}_1(0)}}$ can be figured out to be: $\frac{4\Lambda^2}{\mathcal{T}_0}\frac{4k_BT}{(2\pi)^2\mathcal{T}_0\Lambda^2}\sqrt{\frac{\mathcal{T}_0}{\mathcal{T}_1(0)}}$. The second one $\Lambda^4\sqrt{\frac{\mathcal{T}_0}{\mathcal{T}_1(0)}}\int_0^{2\pi}\frac{d\theta}{(2\pi)^2}\frac{(\sqrt{\frac{\mathcal{T}_0}{\mathcal{T}_1}}\cos\theta-\sin\theta)^2(1+f_B(\xi_{0\Lambda}))}{\xi_{0\Lambda}(1+\frac{\mathcal{T}_0}{\mathcal{T}_1}\cos^2\theta+\frac{\mathcal{T}_1}{\mathcal{T}_0}\sin^2\theta)}$ can be proved to be $\frac{\Lambda^2}{\pi\mathcal{T}_0}\frac{\frac{\mathcal{T}_0}{\mathcal{T}_1(0)}}{\frac{\mathcal{T}_0}{\mathcal{T}_1(0)}+1}(1+f_B(\xi_{0\Lambda}))\sqrt{\frac{\mathcal{T}_0}{\mathcal{T}_1(0)}}$. Combined with these two results, the expression of $A$ is given by:
		\begin{align}
			A&=\frac{\Lambda^2}{\pi\mathcal{T}_0}\sqrt{\frac{\mathcal{T}_0}{\mathcal{T}_1(0)}}(\frac{4k_BT}{\pi\mathcal{T}_0\Lambda^2}+\frac{\frac{\mathcal{T}_0}{\mathcal{T}_1}}{\frac{\mathcal{T}_0}{\mathcal{T}_1}+1}(1+f_B(\xi_{0\Lambda})))\nonumber\\
			&\approx\frac{\Lambda^2}{\pi\mathcal{T}_0}\sqrt{\frac{\mathcal{T}_0}{\mathcal{T}_1(0)}}(\frac{4k_BT}{\pi\mathcal{T}_0\Lambda^2}+1+f_B(\xi_{0\Lambda}))>0\,,
		\end{align} 
		
		%
		
%
	\end{document}